\documentclass[aps,pre,twocolumn,superscriptaddress,10pt,floatfix,longbibliography]{revtex4-2}
\usepackage[colorlinks=true,linkcolor=black,citecolor=black,urlcolor=black]{hyperref}
\usepackage{url}

\usepackage{graphicx}
\usepackage{amsmath,amssymb}
\usepackage{mathtools}
\usepackage{bm}
\usepackage[caption=false]{subfig} 
\usepackage{xcolor}
\colorlet{BLUE}{blue}
\colorlet{MAGENTA}{magenta}
\definecolor{LIGHTBLUE}{RGB}{0,145,200}
\definecolor{GREEN}{RGB}{0,128,0}

\begin{document}

\title{Equilibrium Thermodynamics of  Non-Hermitian Dirac Fermions:\\ Caloric and Magnetic Responses}

\author{Francisco J. Pe\~na}
\thanks{These authors contributed equally to this work.}
\affiliation{Facultad de Ingenier\'ia, Universidad San Sebasti\'an, Lago Panguipulli 1390, Puerto Montt, Chile}

\author{Bastian Castorene}
\thanks{These authors contributed equally to this work.}
\affiliation{Departamento de F\'isica, Universidad T\'ecnica Federico Santa Mar\'ia, Casilla 110, Valpara\'iso, Chile}
\affiliation{Instituto de F\'isica, Pontificia Universidad Cat\'olica de Valpara\'iso, Avenida Universidad 331, Curauma, Valpara\'iso, Chile}

\author{Juan Pablo Esparza}
\thanks{These authors contributed equally to this work.}
\affiliation{Departamento de F\'isica, Universidad T\'ecnica Federico Santa Mar\'ia, Casilla 110, Valpara\'iso, Chile}
\affiliation{Instituto de F\'isica, Pontificia Universidad Cat\'olica de Valpara\'iso, Avenida Universidad 331, Curauma, Valpara\'iso, Chile}

\author{Vladimir Juri\v ci\'c}
\thanks{Corresponding author: vladimir.juricic@usm.cl}
\affiliation{Departamento de F\'isica, Universidad T\'ecnica Federico Santa Mar\'ia, Casilla 110, Valpara\'iso, Chile}

\author{Patricio Vargas}
\affiliation{Departamento de F\'isica, Universidad T\'ecnica Federico Santa Mar\'ia, Casilla 110, Valpara\'iso, Chile}

\date{\today}

\begin{abstract}

We establish scaling relations governing the equilibrium thermodynamics of real-spectrum non-Hermitian Dirac fermions in a magnetic field. Assuming thermalization with respect to the
quasi-Hermitian Hamiltonian, a similarity transformation maps the system at
the same applied field onto a Hermitian Dirac model with reduced velocity,
while the Landau-level spectrum also admits a representation in terms of a
reduced effective magnetic field. This structure yields scaling relations for
the chemical potential, entropy, heat capacities, and orbital
magnetic response in different thermodynamic ensembles. At fixed projected
filling factor (PFF), the self-consistent chemical potential follows the
compressed ladder of Landau levels, and the canonical thermodynamic functions
are rescaled Hermitian responses. At fixed chemical potential, Landau-level crossings generate oscillatory caloric and magnetic responses governed by the same spectral compression. Quasistatic non-Hermitian
deformation at fixed PFF further yields  adiabatic
temperature scaling. More broadly, these results establish a thermodynamic framework for real-spectrum non-Hermitian quantum matter and provide a starting point for incorporating the effects of interactions and disorder within the same formalism.
\end{abstract}
\maketitle

\section{Introduction}
\label{sec:intro}

\begingroup

Non-Hermitian (NH) Hamiltonians provide an effective framework for wave and
quasiparticle systems subject to gain--loss imbalance
\cite{Savoia2016NH-induced,Takata2018Photonic,ElGanainy2018NatPhys,
Liu2020GainAndLoss,Xue2020NHDirac,Cornelius2022Spectral,Li2022Gain-Loss,
Jiang2024TunableNHSE,Peng2024AcousticNHDSMs,Shen2024GainAndLoss}, leakage
\cite{Koutserimpas2018Nonreciprocal,Roccati2022NHPhysics,Li2023Loss-induced},
nonreciprocity
\cite{Hatano1996Localization,Hatano1997Vortex,Hatano1998NHDelocalization,
Ezawa2019NHBoundary,Ghaemi-Dizicheh2023Transport,Reisenbauer2024NHDynamics,
Jana2025Harnessing}, and selective coupling to external reservoirs
\cite{Chaduteau2026Lindbladian}. These settings are encompassed by a broad NH
framework
\cite{BenderBoettcher1998PRL,Ashida2020AdvPhys,Bergholtz2021RMP,
Kawabata2019PRX,Torres2019JPhysMater} and support exceptional degeneracies
\cite{Heiss2004EPsofNHoperators,Muller2008EPsinOPQS,
Heiss2012ThephysicsofEPs,Okugawa2019Topological,Mohammad-Ali2019EPs,
Mandal2021Symmetry,Wu2025Experimental,Yoshida2026HopfEPs,Li2026exceptional},
NH topological phases
\cite{ShenZhenFu2018PRL,Gong2018PRX,Kawabata2019Topological,
Kawabata2019Symmetry,Kawabata2019Classification,Kotz2023Topological,
Xiao2026Symmetry}, and skin effects
\cite{Martinez-Alvarez2018NHRobust,Yao2018Edge,Kunst2018PRL,Song2019NHSE,
Kawabata2020HONHSE,Okuma2020TopologicalNHSE,Zhang2021Observation,
Li2022DynamicNHSE,Liang2022DynamicSignaturesofNHSE,Zhang2022Universal,
Li2024ObservationofDNHSE,Yoshida2024NHMSE}.

Much of NH physics addresses dynamical and nonequilibrium phenomena. A
complementary setting arises for a stationary deformation with an entirely
real spectrum below a threshold
\cite{Mostafazadeh2002PsHI,Mostafazadeh2002PsHII}. When the Hamiltonian is
quasi-Hermitian, a positive-definite metric and a nonsingular similarity
transformation relate it to a Hermitian representative, providing a consistent
basis for equilibrium statistical mechanics
\cite{Mostafazadeh2010IJGMMP,GardasDeffner2016SciRep,Bebiano2020JMP}. A
stability-based equilibrium criterion formulated independently of a
pseudo-Hermitian metric leads to the same stationary real-spectrum setting
\cite{CaoKou2023PRResearch}. Throughout this work, equilibrium means that the
system thermalizes with respect to the quasi-Hermitian Hamiltonian, or
equivalently its Hermitian representative. In an open-system implementation,
the bath or reservoir dynamics must support the corresponding Gibbs ensemble
\cite{Roccati2022NHPhysics,Chaduteau2026Lindbladian,
GoriniKossakowskiSudarshan1976JMP,Lindblad1976CMP}.

We study the thermodynamic consequences of this structure for two-dimensional
Dirac fermions, using graphene as a representative platform. Near charge
neutrality, graphene hosts massless Dirac quasiparticles
\cite{CastroNeto2009RMP,Abergel2010AdvPhys}. A
perpendicular magnetic field reorganizes the spectrum into relativistic Landau
levels, underlying the anomalous integer quantum Hall effect and characteristic
orbital magnetic responses
\cite{Novoselov2005Nature,Zhang2005Nature,Goerbig2011RMP,
ZhengAndo2002PRB,GusyninSharapov2005PRL,GusyninSharapov2006PRB,
McClure1956PR,McClure1960PR,KoshinoAndo2007PRB}. In the present platform,
Landau quantization converts the smooth Dirac continuum into discrete,
field-tunable levels with millielectron-volt spacings. It therefore places the
electronic entropy, heat capacity, orbital response, and caloric effects on
cryogenic scales accessible to thermal measurements in graphene
\cite{FongSchwab2012PRX,Betz2012PRL}.

We consider a minimal NH deformation that preserves the real spectrum for
$|\beta|<1$ and renormalizes the Dirac velocity according to
\cite{JuricicRoy2024CommunPhys}
\begin{equation}
 v_F(\beta)=v\sqrt{1-\beta^2}.
\label{eq:vFbeta}
\end{equation}
All nonzero Landau levels are consequently compressed by the common factor
$\lambda=\sqrt{1-\beta^2}$, while the zero mode and the ordering of the Landau levels are preserved. Since relativistic Landau energies scale as $\sqrt{B}$,
the same spectrum can be represented by the effective field
$B_{\mathrm{eff}}=\lambda^2B$. This effective field parametrizes the spectral
compression, whereas the physical orbital degeneracy remains determined by
the applied field $B$. At fixed $B$, the equilibrium spectral thermodynamics
therefore coincides with that of a Hermitian Dirac Hamiltonian with velocity
$v_F(\beta)$, see also App.~\ref{app:pristine}. 

The related study~\cite{EsparzaPenaVargasJuricic2026} examines quantum
capacitance and eigenstate nonorthogonality at fixed $\mu$, with the deformation
generated microscopically by a sublattice hopping imbalance. The present work
instead establishes general scaling relations for the equilibrium caloric and
orbital thermodynamics under fixed projected filling factor (PFF) and fixed chemical
potential. These relations determine the chemical potential, entropy, heat
capacities, and magnetic response from the corresponding Hermitian functions,
while retaining the physical Landau-level degeneracy. The two works therefore
probe complementary consequences of the same spectral compression: spectral
weight and biorthogonal eigenstates in Ref.~\cite{EsparzaPenaVargasJuricic2026},
and ensemble-dependent thermodynamic scaling and the corresponding observables  here.

Real-spectrum NH Dirac systems and NH Landau quantization have
been studied based on graphene and continuum models~\cite{,Regensburger2016,BagarelloHatano2016RSPA,ZhangFranz2020PRL,Xue2020NHDirac}. Their zero-mode structure and interaction-driven magnetic
catalysis have also been established ~\cite{Roy2025PRD,LeongRoy2025AmplifiedMagneticCatalysis}. Interacting non-Hermitian Dirac systems have also been studied within quantum electrodynamics and Gross-Neveu-Yukawa field theories~\cite{MurshedRoy2024JHEP,MurshedRoy2025SciPost,LeongRoy2026PRB}. A distinct
construction employs a complex magnetic field and generally produces complex
Landau spectra \cite{MontagOzawa2026NHLL}. The present deformation instead
retains a real applied field and remains in the real-spectrum sector for
$|\beta|<1$, allowing the equilibrium Fermi--Dirac treatment developed below.

Here, $\beta$ controls a stationary spectral deformation that may arise from engineered dissipation, gain--loss imbalance, nonreciprocal couplings, or reservoir-induced self-energies. We focus on the thermodynamic signatures of uniform spectral compression when
individual Landau levels remain spectrally resolved, while field sweeps at fixed particle number, finite linewidth, and perturbations introducing additional energy scales are discussed as natural extensions.

\endgroup

\subsection{Key results}

\begingroup

The uniform spectral compression and its mapping to the effective magnetic
field $B_{\rm eff}$, established in Eqs.~\eqref{eq:similarity_H} and
\eqref{eq:Beff}--\eqref{eq:DOS_effective_field}, lead to the thermodynamic
scaling relations summarized in Eq.~\eqref{eq:main_scaling_relations}. At fixed PFF, the
self-consistent chemical potential and heat capacity follow the same
rescaling as the Landau spectrum, as shown in
Eqs.~\eqref{eq:mu_fixedN_scaling} and
\eqref{eq:CB_temperature_scaling_main}. At fixed chemical potential, the resulting oscillations in the heat capacity, entropy, and
orbital magnetic moment  are shifted in
field according to Eqs.~\eqref{eq:crossing_fields} and
\eqref{eq:inverse_field_period}.

A second key result arises because the applied magnetic field controls both the
Landau-level energies and the number of available orbital states. Consequently, the total orbital moment of the projected electron sector and
the corresponding moment per particle scale differently, as given by
Eqs.~\eqref{eq:Mphys_effective_field} and
\eqref{eq:mphys_effective_field}. Quasistatic tuning of $\beta$
is accompanied by reversible work associated with changing the Landau-level
spacing, together with entropy and adiabatic temperature responses. These
effects are summarized in Eq.~\eqref{eq:generalized_force_exact},
Eqs.~\eqref{eq:deltaS_fixedN_Beff} and
\eqref{eq:entropy_integral_identity}, and
Eqs.~\eqref{eq:adiabatic_scaling}--\eqref{eq:differential_caloric_coefficient},
respectively.Together, these scaling relations determine the caloric and magnetic responses
induced by uniform quasi-Hermitian spectral compression at fixed PFF and fixed chemical potential.
\endgroup

\subsection{Organization}

The paper is organized as follows.In Sec.~\ref{sec:model_LL}, we introduce the NH Dirac model, its
quasi-Hermitian mapping, and the relation between its real Landau spectrum and
that of the Hermitian system at the effective magnetic field.
Section~\ref{sec:thermo_framework} develops the thermodynamic framework and
defines the fixed PFF and fixed chemical potential
protocols.
In Sec.~\ref{sec:LL_response}, we analyze the corresponding heat-capacity,
entropy, and orbital magnetic responses, followed by the deformation-induced
caloric response in Sec.~\ref{sec:caloric}.
Section~\ref{sec:discussion_conclusions} summarizes the main results.
Technical details are deferred to the appendices.

\section{Model and quasi-Hermitian mapping}
\label{sec:model_LL}

This section establishes the spectral structure underlying the thermodynamic
analysis. We first construct the quasi-Hermitian form of the
non-Hermitian Dirac Hamiltonian in the real-spectrum interval $|\beta|<1$ and
then derive its Landau-level spectrum and effective-field parametrization,
carefully distinguishing spectral compression from the physical orbital
degeneracy set by the applied magnetic field.

\subsection{Quasi-Hermitian Dirac Hamiltonian}

\begingroup

We consider one spin and valley sector of monolayer graphene in a uniform
perpendicular field $\mathbf B=B\hat z$.  For an electron of charge $-e$,
$\Pi_i=p_i+eA_i$ and $[\Pi_x,\Pi_y]=-i\hbar eB$.  The minimal NH Dirac
Hamiltonian is~\cite{JuricicRoy2024CommunPhys}
\begin{equation}
H_\beta=
 v\sum_{i=x,y}\left(\Gamma_i+\beta M\Gamma_i\right)\Pi_i,
\label{eq:H_NH}
\end{equation}
where $\{\Gamma_i,\Gamma_j\}=2\delta_{ij}$,
$\{M,\Gamma_i\}=0$, and $M^2=1$.  One may choose
$\Gamma_x=\sigma_x$, $\Gamma_y=\sigma_y$, and $M=\sigma_z$.
Because $M\Gamma_i$ is anti-Hermitian, $H_\beta^\dagger\ne H_\beta$ for
$\beta\ne0$.

Equation~\eqref{eq:H_NH} may equivalently be written as
$H_\beta=(1+\beta M)H_0$, where $H_0=v \Gamma_i \Pi_i$ is the standard Hermitian Hamiltonian for massless Dirac fermions in the magnetic field. Consequently, the NH deformation
introduces a single spectral parameter, $\lambda=\sqrt{1-\beta^2}$, which  is
the origin of the thermodynamic relations derived below; perturbations
that do not share it need not obey the same scaling.

Let
\begin{equation}
\theta=\operatorname{arctanh}\beta,
\quad {\rm and} \quad  
\mathcal S=e^{\theta M/2}.
\label{eq:similarity_definitions}
\end{equation}
Using $\{M,\Gamma_i\}=0$ gives
\begin{equation}
\mathcal S\Gamma_i\mathcal S^{-1}
=
\cosh\theta\,\Gamma_i+\sinh\theta\,M\Gamma_i,
\label{eq:similarity_gamma}
\end{equation}
and therefore
\begin{equation}
H_\beta=\lambda\,\mathcal S H_0\mathcal S^{-1}.
\label{eq:similarity_H}
\end{equation}
If $H_0|n\rangle=\varepsilon_n|n\rangle$, the corresponding right and left
eigenvectors of $H_\beta$ can be chosen as
\begin{equation}
|\psi_n^{R}\rangle=\mathcal S|n\rangle,
\qquad
\langle\psi_n^{L}|=\langle n|\mathcal S^{-1},
\qquad
E_{n,\beta}=\lambda\varepsilon_n.
\label{eq:biorthogonal_eigenstates}
\end{equation}
They satisfy $\langle\psi_m^{L}|\psi_n^{R}\rangle=\delta_{mn}$. Thus the
spectrum is uniformly compressed while the eigenvectors are transformed
nonunitarily. The thermodynamic observables considered here depend on the former, whereas the operator $\eta$ introduced below defines the consistent
inner product for the latter. The positive-definite metric operator 
\begin{equation}
\eta=\mathcal S^{-2}=e^{-\theta M}
\label{eq:metric_eta}
\end{equation}
satisfies
\begin{equation}
H_\beta^\dagger\eta=\eta H_\beta.
\label{eq:pseudo_hermiticity}
\end{equation}
Equation~\eqref{eq:pseudo_hermiticity} establishes the pseudo-Hermiticity of
$H_\beta$, while 
the positive definiteness of $\eta$ places this NH operator in the
quasi-Hermitian class, ensuring a physical inner product and a nonsingular
mapping to its Hermitian representative~\cite{Scholtz1992AnnPhys,Mostafazadeh2002PsHII,Mostafazadeh2010IJGMMP}.
The equilibrium construction is completed by the Gibbs-operator identity
\begin{equation}
e^{-H_\beta/(k_{\mathrm B}T)}
=
\mathcal S
e^{-\lambda H_0/(k_{\mathrm B}T)}
\mathcal S^{-1},
\label{eq:gibbs_similarity}
\end{equation}
which implies the spectral trace equivalence
\begin{equation}
\operatorname{Tr}e^{-H_\beta/(k_{\mathrm B}T)}
=
\operatorname{Tr}e^{-\lambda H_0/(k_{\mathrm B}T)}.
\label{eq:gibbs_trace}
\end{equation}
\begingroup

The one-particle similarity transformation lifts directly to the fermionic
Fock space. Because the second-quantized transformation
is generated by a number-conserving one-body operator, it commutes with the
total-number operator $\widehat N$. Hence
\(
\widehat H_\beta-\mu\widehat N
=
\lambda\widehat{\mathcal S}
[\widehat H_0-(\mu/\lambda)\widehat N]
\widehat{\mathcal S}^{-1}
\), and at fixed $B$ the grand partition function obeys
\begin{equation}
\Xi_\beta(T,\mu;B)
=
\Xi_0\!\left(\frac{T}{\lambda},
\frac{\mu}{\lambda};B\right).
\label{eq:grand_partition_temperature_scaling}
\end{equation}
Equations~\eqref{eq:gibbs_similarity}--
\eqref{eq:grand_partition_temperature_scaling} are understood either within a
finite-band lattice regularization or after normal ordering and projection
relative to the filled valence background. The similarity transformation
preserves the corresponding regularized trace.
Thus every equilibrium quantity constructed only from the single-particle
eigenvalues is identical to that of the Hermitian Dirac Hamiltonian with the
effective velocity in Eq.~\eqref{eq:vFbeta}. This construction assumes
thermalization with respect to $\widehat H_\beta$, or, equivalently, to its Hermitian
representative. If $H_\beta$ is obtained as an effective no-jump or reduced open-system
generator, compatibility of the complete bath and jump dynamics with this
Gibbs state is an additional physical
requirement~\cite{Minganti2019PRA,Davies1974CMP,Kossakowski1977CMP}.
Equation~\eqref{eq:grand_partition_temperature_scaling} shows the same
scaling at fixed applied field, with both temperature and chemical potential
rescaled by the spectral-compression factor. The equivalent
 representation in terms of an effective magnetic field, which is introduced below,  keeps $T$ and $\mu$ unchanged
and instead maps $B$ to $B_{\mathrm{eff}}$. At $|\beta|=1$, the metric and similarity transformation become singular, and
the quasi-Hermitian description ceases to apply.
\endgroup
\endgroup

\subsection{Landau spectrum and effective-field mapping}

\begingroup

The deformed matrices
$\alpha_i(\beta)=\Gamma_i+\beta M\Gamma_i$ satisfy
\begin{equation}
\{\alpha_i(\beta),\alpha_j(\beta)\}
=2(1-\beta^2)\delta_{ij},
\label{eq:alpha_clifford}
\end{equation}
so the standard ladder-operator construction survives with the velocity in
Eq.~\eqref{eq:vFbeta}.  The cyclotron scale is
\begin{equation}
\omega_c^{(1)}(\beta)
=
\sqrt{\frac{2eBv_F^2(\beta)}{\hbar}},
\label{eq:omegac}
\end{equation}
and the nonzero Landau levels are
\begin{equation}
E_{n,\pm}(B,\beta)
=
\pm\hbar\omega_c^{(1)}(\beta)\sqrt n,
\qquad n=1,2,\ldots,
\label{eq:LLs_nonzero}
\end{equation}
together with the graphene zero mode $E_0=0$.  Equivalently,
\begin{equation}
E_{n,s}(B,\beta)
=
s\lambda v\sqrt{2e\hbar B|n|},
\label{eq:LLs}
\end{equation}
where $s=\pm$ for $n\ge1$ and the zero level is counted once. The zero
mode remains pinned at the charge-neutrality energy because the deformation
rescales the full Dirac operator without introducing an additive mass or
energy offset. The adjacent positive-energy spacing is
\begin{equation}
\Delta_n(\beta)
=E_{n+1,+}-E_{n,+}
=\lambda v\sqrt{2e\hbar B}\,
\bigl(\sqrt{n+1}-\sqrt n\bigr),
\label{eq:LL_spacing}
\end{equation}
so every thermally activated Landau-level scale carries the same factor
$\lambda$.

Since the relativistic levels scale as $\sqrt B$, define
\begin{equation}
B_{\mathrm{eff}}
=
\lambda^2B
=
(1-\beta^2)B.
\label{eq:Beff}
\end{equation}
Then
\begin{equation}
E_{n,s}(B,\beta)=E_{n,s}(B_{\mathrm{eff}},0),
\label{eq:LL_effective_field}
\end{equation}
and the reduced density of states obeys
\begin{equation}
D_\beta(E;B)=D_0(E;B_{\mathrm{eff}}).
\label{eq:DOS_effective_field}
\end{equation}
This is the master spectral identity. It preserves the Landau-level index,
the zero mode, and the ordering of the positive- and negative-energy branches.
The mapping should nevertheless be interpreted spectrally: it does not replace
the physical field in the orbital degeneracy. For a sample of area $A$,
$\mathcal D_B=eBA/h$, whereas
$\mathcal D_{B_{\mathrm{eff}}}=\lambda^2\mathcal D_B$. Reduced functions
therefore collapse directly, while physical extensive quantities acquire the
degeneracy factors derived below.
The corresponding scaling of physical quantities is derived in
Appendix~\ref{app:scaling}.

The two representations emphasize different aspects of the scaling: the
fixed-field form makes the uniform energy compression explicit through
$T\mapsto T/\lambda$ and $\mu\mapsto\mu/\lambda$, while the effective-field
form is convenient for describing the magnetic-field dependence.
\endgroup

\section{Thermodynamic framework}
\label{sec:thermo_framework}

We now formulate the thermodynamics of the NH Dirac system described by the effective Hamiltonian in Eq.~\eqref{eq:H_NH}. 
We first introduce thermodynamic quantities normalized by the orbital
degeneracy and relate them to their physical extensive counterparts. We then
define the thermodynamic constraints considered throughout the paper and derive
the orbital magnetic response while retaining the field dependence of the
Landau-level degeneracy.

\subsection{Thermodynamic functions}

\begingroup

Because the spectrum is real, the equilibrium thermodynamics follows directly
from the standard Fermi--Dirac construction for noninteracting fermions
\cite{Callen1985,PathriaBeale2011}. Restricting the spectrum to a set
$\mathcal C$, the particle number and internal energy are
\begin{align}
N_{\mathcal C}
&=
\int_{\mathcal C}dE\,D(E;B,\beta)f(E,\mu,T),
\label{eq:N_DOS}\\
U_{\mathcal C}
&=
\int_{\mathcal C}dE\,E D(E;B,\beta)f(E,\mu,T),
\label{eq:U_DOS}
\end{align}
while the entropy is
\begin{align}
S_{\mathcal C}
&=-k_{\mathrm B}\int_{\mathcal C}dE\,D(E;B,\beta)
\nonumber\\
&\quad\times
\left[f\ln f+(1-f)\ln(1-f)\right],
\label{eq:S_DOS}
\end{align}
where
\begin{equation}
f(E,\mu,T)=
\frac{1}{e^{(E-\mu)/(k_{\mathrm B}T)}+1}.
\label{eq:fermi_function}
\end{equation}
The corresponding zero-field benchmark is presented in
Appendix~\ref{app:pristine}.

In the presence of Landau levels, we focus on the electron sector
\begin{equation}
\mathcal C_{0,+}=\{E_0=0\}\cup\{E_{n,+}>0\},
\end{equation}
and take the negative-energy branch as the filled valence reference. The
thermodynamic and magnetic responses obtained below are therefore defined
relative to the filled valence band.

Each Landau level carries the orbital degeneracy
\begin{equation}
\mathcal D_B=\frac{A}{2\pi\ell_B^2}=\frac{eBA}{h},
\qquad
\ell_B=\sqrt{\frac{\hbar}{eB}},
\label{eq:landau_degeneracy}
\end{equation}
per spin and valley, while the combined spin--valley multiplicity is
$g=g_sg_v=4$. It is then convenient to separate the field-dependent orbital
multiplicity from the Landau-level spectrum by writing
\begin{equation}
D_{0,+}^{\mathrm{phys}}(E;B,\beta)
=
\mathcal D_B D_{0,+}(E;B,\beta),
\label{eq:DOS_LL_physical}
\end{equation}
with
\begin{equation}
D_{0,+}(E;B,\beta)
=
g\left[
\delta(E)+
\sum_{n=1}^{n_{\max}}
\delta\!\left(E-E_{n,+}(B,\beta)\right)
\right].
\label{eq:DOS_LL}
\end{equation}
The corresponding occupation normalized by the orbital degeneracy is
\begin{align}
N_{0,+}(T,\mu;B,\beta)
&=
g\left[
f(0,\mu,T)
+\sum_{n=1}^{n_{\max}}f(E_{n,+},\mu,T)
\right].
\label{eq:N_zero_positive}
\end{align}
Restoring the orbital multiplicity gives the physical particle number and
two-dimensional density,
\begin{align}
N_{0,+}^{\mathrm{phys}}
&=\mathcal D_BN_{0,+},\\
n_{2\mathrm D}
&=\frac{eB}{h}N_{0,+}.
\label{eq:Nphysical_from_reduced}
\end{align}

\begingroup

For the same reason, we introduce thermodynamic functions normalized by
$\mathcal D_B$,
\begin{equation}
(U,S,\Omega)=
\frac{
(U^{\mathrm{phys}},S^{\mathrm{phys}},\Omega^{\mathrm{phys}})
}{
\mathcal D_B
}.
\label{eq:reduced_extensive_quantities}
\end{equation}
This normalization separates the spectral response from the field-dependent
number of orbital states. The physical extensive quantities are recovered by
restoring $\mathcal D_B$, which must be done before taking magnetic-field
derivatives.
\endgroup

The zero mode is counted once, equivalently by taking the lower integration
limit as $0^-$. It contributes to $N_{0,+}$, $S$, and $\Omega$. At finite temperature, only Landau levels within an energy range of order
$k_{\mathrm B}T$ around the chemical potential contribute appreciably to the
occupation, entropy, and heat capacity. We therefore truncate the Landau-level
sum at $n_{\max}$ such that
$E_{n_{\max},+}-\mu \gg k_{\mathrm B}T$. Levels above this cutoff have
exponentially small Fermi occupation and make a negligible contribution to the
thermodynamic quantities. Convergence is verified by increasing $n_{\max}$ and
confirming that the results remain unchanged.

\endgroup
\subsection{Thermodynamic constraints}

\begingroup

The thermodynamic response depends on the constraint imposed as $T$ or $B$ is
varied. We consider two protocols: fixed PFF and fixed
chemical potential. Since
$N_{0,+}^{\mathrm{phys}}=\mathcal D_B N_{0,+}$ and
$\mathcal D_B=BA/\Phi_0$ counts the magnetic flux quanta through the sample,
$N_{0,+}$ gives the number of projected electrons per flux quantum. The
fixed-filling constraint is therefore
\begin{equation}
N_{0,+}(T,\mu;B,\beta)=N_{\rm c},
\label{eq:Nconstraint}
\end{equation}
with the chemical potential determined self-consistently.

At fixed $B$, the orbital degeneracy is constant, and fixing $N_{\rm c}$ also
fixes the physical projected particle number. During a field sweep,
$\mathcal D_B\propto B$, so that
$N_{0,+}^{\mathrm{phys}}=\mathcal D_BN_{\rm c}$ varies with the applied
field. A fixed-$N_{\rm c}$ sweep therefore preserves the projected filling
factor, while the physical particle number and density follow the changing
orbital degeneracy.

The canonical heat capacity along this constrained trajectory is
\begin{equation}
C_B(T)=
\left(\frac{\partial U}{\partial T}\right)_{B,N_{\rm c},\beta}.
\label{eq:C}
\end{equation}
Because $\mu$ varies with temperature to maintain
Eq.~\eqref{eq:Nconstraint}, the derivative includes its implicit temperature
dependence. Defining
\begin{equation*}
N_T=
\left(\frac{\partial N_{0,+}}{\partial T}\right)_{B,\mu,\beta},
\qquad
N_\mu=
\left(\frac{\partial N_{0,+}}{\partial\mu}\right)_{B,T,\beta},
\end{equation*}
and introducing $U_T$ and $U_\mu$ analogously, one obtains
\begin{equation}
\begin{aligned}
\left(\frac{d\mu}{dT}\right)_{B,N_{\rm c},\beta}
&=-\frac{N_T}{N_\mu},\\
C_B
&=U_T-U_\mu\frac{N_T}{N_\mu}.
\end{aligned}
\label{eq:canonical_chain_rule}
\end{equation}
The second term incorporates the self-consistent shift of the chemical
potential as the thermal population is redistributed among the Landau levels.

In the grand-canonical protocol, the reservoir fixes $\mu$, while
$N_{0,+}$ varies with $B$, $T$, and $\beta$. We characterize the thermal
response per projected electron by
\begin{align}
c_\mu(B,T)
&=\frac{T}{N_{0,+}}
\left(\frac{\partial S}{\partial T}\right)_{B,\mu,\beta}
\label{eq:c_mu}\\
&=\frac{1}{N_{0,+}}
\left[
\left(\frac{\partial U}{\partial T}\right)_{B,\mu,\beta}
-\mu
\left(\frac{\partial N_{0,+}}{\partial T}\right)_{B,\mu,\beta}
\right].
\label{eq:c_mu_equiv}
\end{align}
Both the entropy response and the occupation are evaluated at the same
thermodynamic state. The term proportional to
$\mu(\partial N_{0,+}/\partial T)_{B,\mu,\beta}$ incorporates the energy
associated with particle exchange with the reservoir. For later use, we write
\begin{equation}
N(B)\equiv N_{0,+}(T,\mu;B,\beta).
\label{eq:N_of_B_definition}
\end{equation}

A field sweep at fixed physical projected particle number
$N_{\rm phys}^{\star}$ would impose
\begin{equation}
N_{0,+}(T,\mu;B,\beta)
=
\frac{N_{\rm phys}^{\star}}{\mathcal D_B}
=
\frac{hN_{\rm phys}^{\star}}{eAB}.
\label{eq:fixed_physical_constraint}
\end{equation}
In this protocol, the PFF decreases as $1/B$ and
compensates for the increasing orbital degeneracy. The present analysis
focuses on fixed $N_{\rm c}$ and fixed $\mu$, while the fixed-number protocol
provides a natural extension of the ensemble framework.

For $N_{\rm c}=75$, Eq.~\eqref{eq:Nphysical_from_reduced} gives
\begin{equation*}
n_{2\mathrm D}
=
1.81\times10^{12}
\left(\frac{B}{1~\mathrm T}\right)
\mathrm{cm}^{-2}.
\end{equation*}
The physical density therefore varies linearly with $B$ along a
fixed-$N_{\rm c}$ field sweep. In the following, ``canonical'' refers to
temperature derivatives evaluated at fixed $B$, $N_{\rm c}$, and $\beta$.

\endgroup

\subsection{Orbital magnetic response}

\begingroup

The orbital magnetic response requires particular care because the applied
field controls both the Landau-level energies and the number of available
orbital states. We begin with the grand potential normalized by the orbital
degeneracy,
\begin{align}
\Omega(T,\mu;B,\beta)
&=-k_{\mathrm B}T
\int_{\mathcal C_{0,+}}dE\,D_{0,+}(E;B,\beta)
\nonumber\\
&\quad\times
\ln\!\left[1+e^{-(E-\mu)/(k_{\mathrm B}T)}\right].
\label{eq:Omega_DOS}
\end{align}
Since $\Omega^{\mathrm{phys}}=\mathcal D_B\Omega$, the magnetic response
contains contributions from both the Landau-level dispersion and the
field-dependent orbital multiplicity $\mathcal D_B\propto B$.

The total orbital magnetic moment of the projected electron sector is
therefore
\begin{align}
M_{0,+}^{\mathrm{phys}}
&=-\left(
\frac{\partial\Omega^{\mathrm{phys}}}{\partial B}
\right)_{T,\mu,\beta}
\nonumber\\
&=\mathcal D_BM_{\mathrm{sp}}
-\left(\frac{\partial\mathcal D_B}{\partial B}\right)\Omega,
\label{eq:Mphysical_from_reduced}
\end{align}
where
\begin{equation}
M_{\mathrm{sp}}
=
-\left(\frac{\partial\Omega}{\partial B}\right)_{T,\mu,\beta}
\label{eq:M}
\end{equation}
is the spectral contribution normalized by the orbital degeneracy.

\endgroup
\section{Thermodynamics in a magnetic field}
\label{sec:LL_response}

We now translate the spectral mapping into the corresponding thermodynamic relations. The
uniform compression of the ladder of Landau levels organizes the response in both
ensembles considered below: at fixed PFF, the chemical
potential follows the spectrum self-consistently, whereas at fixed chemical
potential the reservoir provides an external energy reference and the
compression is revealed through displaced Landau-level crossings.

\subsection{Scaling relations}

The effective-field representation
\[
B_{\mathrm{eff}}=\lambda^{2}B=(1-\beta^{2})B
\]
provides the common structure of the caloric and magnetic responses.
As shown in Appendix~\ref{app:scaling}, it yields the scaling relations
\begin{equation}
\begin{aligned}
\mu_{\beta}(T,B;N_{\rm c})
&=\mu_{0}(T,B_{\mathrm{eff}};N_{\rm c}),\\
C_{B,\beta}(T,B;N_{\rm c})
&=C_{B,0}(T,B_{\mathrm{eff}};N_{\rm c}),\\
c_{\mu,\beta}(T,\mu;B)
&=c_{\mu,0}(T,\mu;B_{\mathrm{eff}}),\\
M_{0,+,\beta}^{\mathrm{phys}}(T,\mu;B)
&=M_{0,+,0}^{\mathrm{phys}}
(T,\mu;B_{\mathrm{eff}}),\\
m_{0,+,\beta}^{\mathrm{phys}}(T,\mu;B)
&=\lambda^{2}
m_{0,+,0}^{\mathrm{phys}}
(T,\mu;B_{\mathrm{eff}}).
\end{aligned}
\label{eq:main_scaling_relations}
\end{equation}
These identities express the complete thermodynamic response in terms of the
Hermitian problem. At fixed $B$, the compression may be viewed as the energy
rescaling $T\mapsto T/\lambda$ and $\mu\mapsto\mu/\lambda$; at fixed $T$ and
$\mu$, it appears as the field transformation $B\mapsto B_{\mathrm{eff}}$.
The most convenient representation depends on the ensemble constraint. For fixed
$N_{\rm c}$, the self-consistent chemical potential follows the compressed
spectrum, while for fixed $\mu$ the same compression shifts the sequence of
Landau-level crossings. The magnetic relations additionally retain the
field-dependent orbital degeneracy, which distinguishes the total orbital
moment from the response per projected electron.

For the numerical results, the Landau-level sums are evaluated from
Eqs.~\eqref{eq:LLs_nonzero} and~\eqref{eq:LLs} using
$v=1.0\times10^{6}~\mathrm{m\,s^{-1}}$. At each parameter point, the cutoff is
increased until $E_{n_{\max},+}-|\mu|\geq20k_{\mathrm B}T$, after which all
reported observables are converged to a relative tolerance of $10^{-8}$. The
fixed-$N_{\rm c}$ chemical potential satisfies Eq.~\eqref{eq:Nconstraint}
with an occupation residual below $10^{-10}$. Heat capacities are obtained
from analytic temperature derivatives, including the self-consistent term in
Eq.~\eqref{eq:canonical_chain_rule}, and the spectral magnetic response in
Eq.~\eqref{eq:M} is evaluated using
$\partial_B E_{n,+}=E_{n,+}/(2B)$ at fixed $\beta$.

\subsection{Fixed projected filling factor}
\label{subsec:fixed_Nc}

\begingroup

We first fix the PFF at
$N_{0,+}=N_{\rm c}=75$. The chemical potential is then determined at every
$(T,B,\beta)$ from
\begin{equation}
\mu=\mu(T,B;\beta,N_{\rm c}),
\label{eq:mu_selfconsistent}
\end{equation}
and any observable is evaluated along the resulting thermodynamic trajectory,
\begin{equation}
X_{N_{\rm c}}
=X[T,\mu(T,B;\beta,N_{\rm c});B,\beta].
\label{eq:X_fixedN_map}
\end{equation}
Equivalently, at fixed applied field, the chemical potential scales as
\begin{equation}
\mu_\beta(T,B;N_{\rm c})
=\lambda\mu_0(T/\lambda,B;N_{\rm c}).
\label{eq:mu_fixedN_scaling}
\end{equation}
Thus, the chemical potential follows the uniform compression of
the Landau levels  according to Eq.~\eqref{eq:mu_fixedN_scaling}.

The spectral compression also controls the canonical heat capacity. At fixed
$B$, one finds
\begin{equation}
C_{B,\beta}(T,B;N_{\rm c})
=C_{B,0}(T/\lambda,B;N_{\rm c}),
\label{eq:CB_temperature_scaling_main}
\end{equation}
so increasing $\beta$ moves the thermally activated structure toward lower
physical temperatures while preserving the Hermitian line shape.
Figure~\ref{fig:CB_fixedN_75} shows this response normalized by
$N_{\rm c}k_{\mathrm B}$ and directly illustrates the temperature rescaling
of the canonical heat capacity.

\begin{figure}[t!]
\centering
\includegraphics[width=\linewidth]{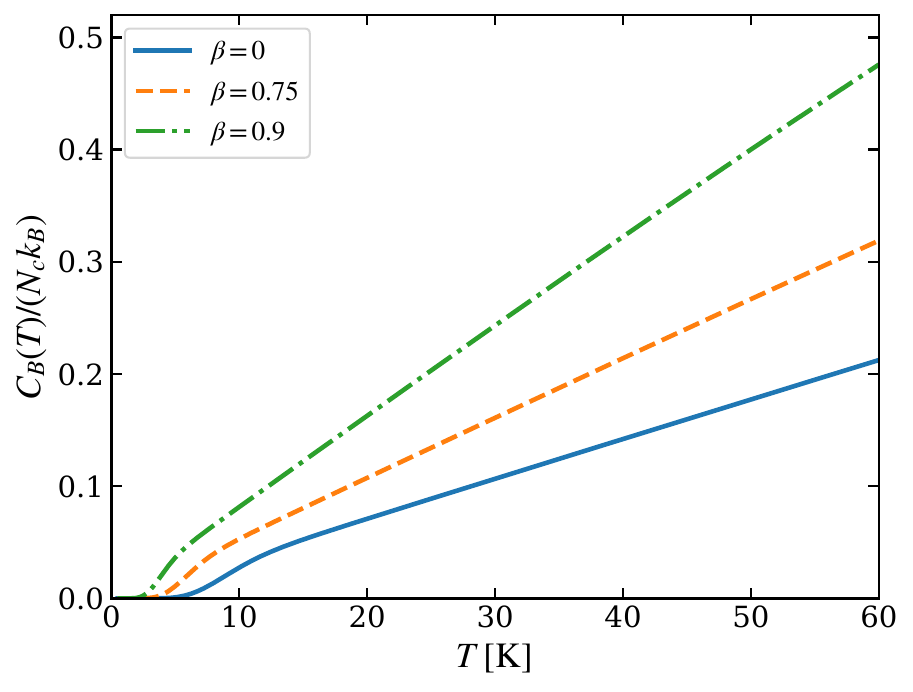}
\caption{Canonical heat capacity per electron at
$B=1~\mathrm T$ and PFF $N_{\rm c}=75$. Spectral
compression shifts the Hermitian response toward lower temperatures. According
to Eq.~\eqref{eq:CB_temperature_scaling_main}, all curves represent the same
canonical function evaluated at the rescaled temperature $T/\lambda$.}
\label{fig:CB_fixedN_75}
\end{figure}

\endgroup

\subsection{Fixed chemical potential}
\label{subsec:fixed_mu}

\begin{figure}[t!]
\centering
\includegraphics[width=\linewidth]{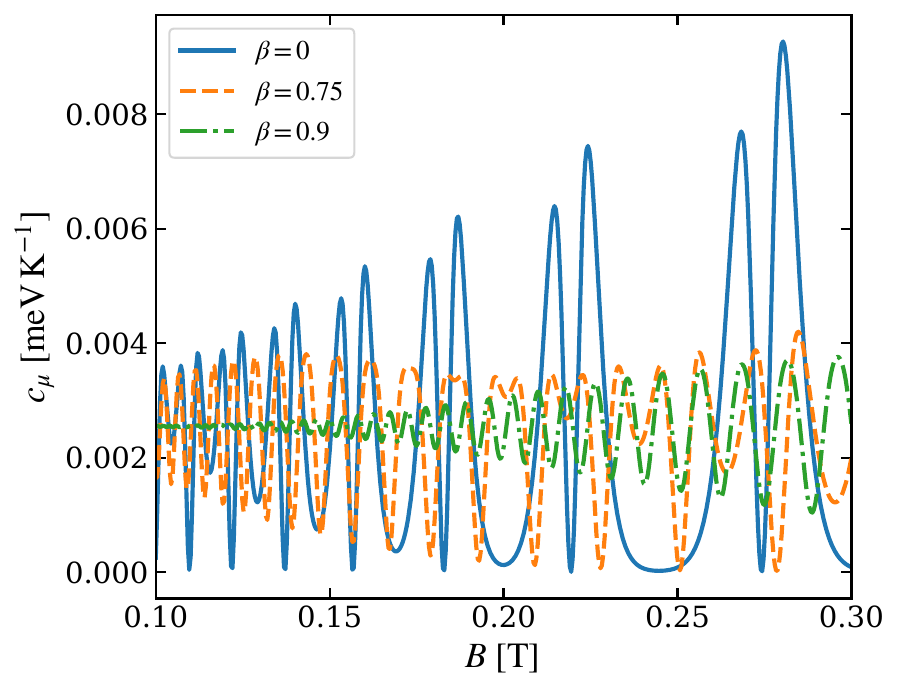}
\caption{Grand-canonical heat capacity per electron at $T=2~\mathrm K$ and $\mu=38~\mathrm{meV}$. Successive
Landau-level crossings generate oscillatory, thermally broadened structures.
For an isolated level, the fixed-$\mu$ heat-capacity contribution vanishes at
the crossing $E_{n,+}(B,\beta)=\mu$ and develops maxima on either side
within the thermal window. The deformation-dependent field locations of these
structures are controlled by $B_{\mathrm{eff}}=(1-\beta^2)B$ through
Eq.~\eqref{eq:cmu_effective_field}.}
\label{fig:Cmu_fixedmu}
\end{figure}

\begingroup

We now turn to the grand-canonical protocol, with the chemical potential fixed
at $\mu=38~\mathrm{meV}$. As the magnetic field is swept, successive Landau
levels pass through the reservoir energy and generate oscillatory caloric
features. Their deformation dependence is governed by the relation
\begin{equation}
c_{\mu,\beta}(T,\mu;B)
=
c_{\mu,0}(T,\mu;B_{\mathrm{eff}}),
\label{eq:cmu_effective_field}
\end{equation}
so that the non-Hermitian deformation shifts the response 
according to $B_{\mathrm{eff}}=\lambda^2B$.

The crossings $E_{n,+}(B,\beta)=\mu$ occur at
\begin{align}
B_n(\beta)
&=\frac{\mu^2}{2e\hbar v^2\lambda^2n},
\label{eq:crossing_fields}\\
\Delta\!\left(\frac1B\right)
&=\frac{2e\hbar v^2\lambda^2}{\mu^2}.
\label{eq:inverse_field_period}
\end{align}
The inverse-field period therefore scales as $\lambda^2$, while the
oscillation frequency scales as $\lambda^{-2}$. The field positions of the
caloric structures thus directly encode the spectral-compression factor.

Figure~\ref{fig:Cmu_fixedmu} displays the resulting oscillations. For an
individual Landau level, the contribution to $c_\mu$ is proportional to
$(E_{n,+}-\mu)^2f(1-f)$: it vanishes at the  crossing and develops two
thermally broadened peaks within the surrounding Fermi window. Increasing
$\beta$ shifts the full sequence toward larger applied fields according to
$B_n(\beta)\propto\lambda^{-2}$. The peak amplitudes also reflect the
field-dependent projected occupation $N_{0,+}(B)$ entering the normalization
of $c_\mu$.

\endgroup

\begin{figure}[t!]
\centering
\includegraphics[width=\linewidth]{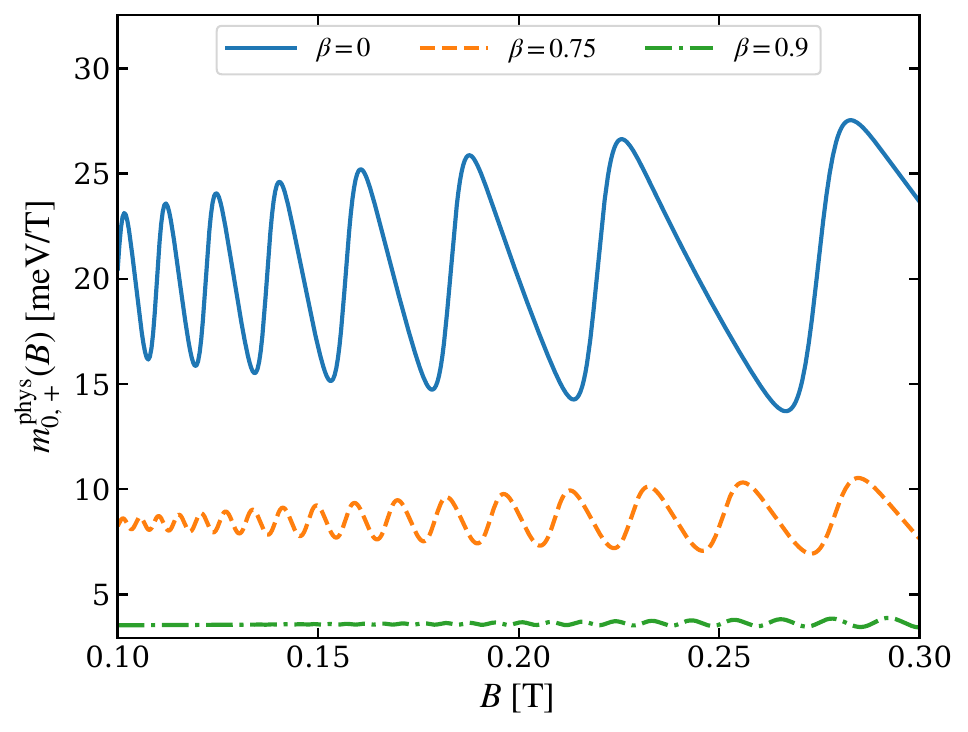}
\caption{Orbital magnetic moment per electron at
$T=2~\mathrm K$ and $\mu=38~\mathrm{meV}$. Its field displacement
and deformation-dependent amplitude follow
$m_{0,+,\beta}^{\mathrm{phys}}(B)=\lambda^2
m_{0,+,0}^{\mathrm{phys}}(B_{\mathrm{eff}})$. The calculation excludes the
regularized negative-energy Dirac-sea contribution.}
\label{fig:m_fixedmu}
\end{figure}

\begingroup

The orbital response in the fixed-$\mu$ protocol inherits the same
 structure in terms of the effective field. The spectral contribution, the total orbital
moment of the electron in the projected sector, and the corresponding moment per particle obey
\begin{align}
M_{\mathrm{sp},\beta}(B)
&=\lambda^2M_{\mathrm{sp},0}(B_{\mathrm{eff}}),
\label{eq:Msp_effective_field}\\
M_{0,+,\beta}^{\mathrm{phys}}(B)
&=M_{0,+,0}^{\mathrm{phys}}(B_{\mathrm{eff}}),
\label{eq:Mphys_effective_field}\\
m_{0,+,\beta}^{\mathrm{phys}}(B)
&=\lambda^2m_{0,+,0}^{\mathrm{phys}}(B_{\mathrm{eff}}).
\label{eq:mphys_effective_field}
\end{align}
The field-dependent orbital degeneracy distinguishes the last two relations.
The total orbital moment is displaced along the field axis by
$B_{\mathrm{eff}}=\lambda^2B$, whereas the moment per projected electron also
acquires the amplitude factor $\lambda^2$.

Figure~\ref{fig:m_fixedmu} shows both effects. At matched
$B_{\mathrm{eff}}$, the rescaled quantity
$m_{0,+,\beta}^{\mathrm{phys}}/\lambda^2$ collapses onto the Hermitian curve,
while the total orbital moment collapses through the field rescaling alone.
The sign displayed in Fig.~\ref{fig:m_fixedmu} corresponds to the projected
$(0,+)$ electron sector, with the filled valence band taken as the reference.
\endgroup

\section{Caloric response to spectral compression}
\label{sec:caloric}

Within the effective description, we consider quasistatic
protocols connecting stationary real-spectrum Hamiltonians with different
values of $\beta$. Changing the spectral compression is then associated with
reversible work, entropy changes, and adiabatic temperature variations. We
first establish the general response relations and then apply them to fixed
PFF and fixed chemical potential.

\subsection{Quasistatic work and response coefficients}

\begingroup

We regard $\beta$ as a quasistatically tunable parameter of the stationary
Hamiltonian. Its conjugate thermodynamic response is defined in the
grand-canonical and canonical ensembles by
\begin{equation}
X_\beta^{(\mu)}
=-\left(\frac{\partial\Omega}{\partial\beta}\right)_{T,\mu,B},
\qquad
X_\beta^{(N)}
=-\left(\frac{\partial F}{\partial\beta}\right)_{T,N,B},
\label{eq:generalized_beta_forces}
\end{equation}
where $F=U-TS$. Here $N$ denotes the PFF; for the
fixed-filling protocol considered below, $N=N_{\rm c}$. The corresponding
physical projected particle number is
$N_{0,+}^{\mathrm{phys}}=\mathcal D_BN$.

Because every nonzero energy scales with
$\lambda=\sqrt{1-\beta^2}$, the response to the deformation is fixed directly
by the internal energy,
\begin{equation}
X_\beta=\frac{\beta}{\lambda^2}U_\beta.
\label{eq:generalized_force_exact}
\end{equation}
For a quasistatic variation of $\beta$ at fixed applied magnetic field, the
change in spectral compression contributes to the reduced thermodynamic
potentials according to
\begin{align}
dU&=T\,dS+\mu\,dN-X_\beta\,d\beta,
\label{eq:first_law_beta}\\
dF&=-S\,dT+\mu\,dN-X_\beta^{(N)}\,d\beta,
\label{eq:dF_beta}\\
d\Omega&=-S\,dT-N\,d\mu-X_\beta^{(\mu)}\,d\beta.
\label{eq:dOmega_beta}
\end{align}
The term $-X_\beta d\beta$ is the reversible work contribution
associated with changing the spectral compression.

The response of $X_\beta^{(N)}$ to further tuning of $\beta$ defines the
canonical deformation susceptibility,
\begin{equation}
\chi_\beta^{(N)}
\equiv
\left(\frac{\partial X_\beta^{(N)}}{\partial\beta}\right)_{T,N,B}
=
\frac{U_\beta+\beta^2TC_{B,\beta}}{(1-\beta^2)^2}.
\label{eq:deformation_susceptibility}
\end{equation}
For the projected positive-energy sector,
$U_\beta\geq0$ and $C_{B,\beta}\geq0$, yielding a positive susceptibility
whose magnitude increases as the real-spectrum threshold is approached, and eventually diverges as $\beta\to\pm1$. Finite temperature, linewidth, and additional energy scales cut off this ideal divergence before the singular threshold.

\endgroup

\subsection{Fixed projected filling factor}
\label{subsec:caloric_fixedN}

\begin{figure}[t]
\centering
\includegraphics[width=\linewidth]{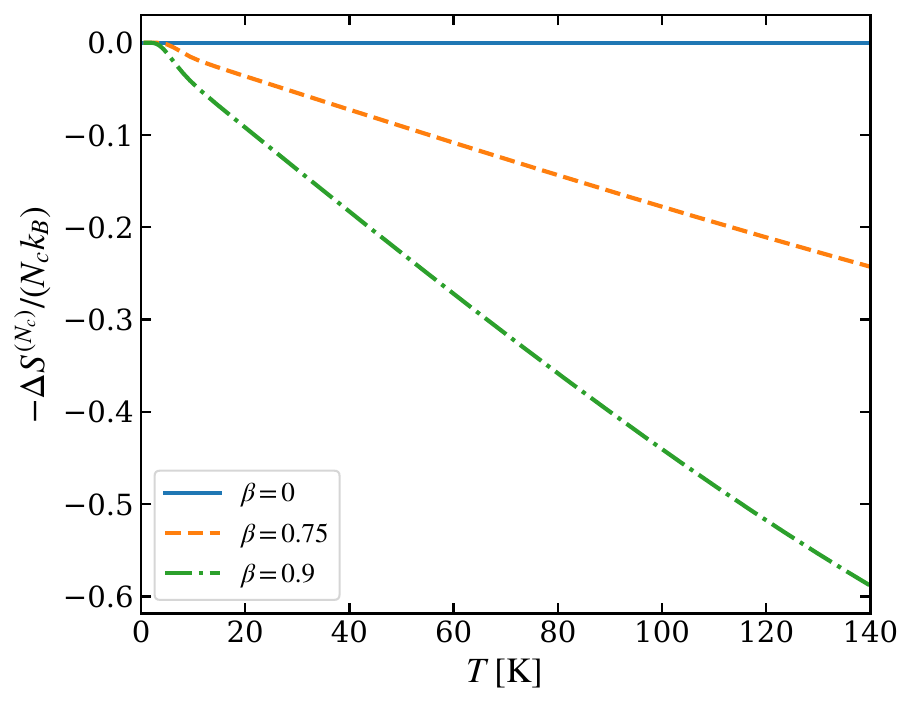}
\caption{Isothermal entropy change at fixed
$N_{\rm c}=75$ and $B=1~\mathrm T$.  Negative values of
$-\Delta S^{(N_{\rm c})}/(N_{\rm c}k_{\mathrm B})$ correspond to entropy
increase under spectral compression.  The residual entropy of the fractionally
filled ideal Landau level is common to both spectra and cancels from the
difference.}
\label{fig:deltaS_fixedN_75}
\end{figure}

\begingroup

We first consider an isothermal change of $\beta$ at fixed $B$ and projected
filling factor $N_{\rm c}$. The entropy change per projected electron is
defined by
\begin{equation}
-\Delta s^{(N_{\rm c})}(T)
=-\frac{S_\beta(T,B;N_{\rm c})-S_0(T,B;N_{\rm c})}
{N_{\rm c}k_{\mathrm B}}.
\label{eq:deltaS_fixedN}
\end{equation}
The  scaling of the canonical entropy gives
\begin{align}
\Delta S_\beta^{(N_{\rm c})}(T,B)
&=S_{0,N_{\rm c}}(T,B_{\mathrm{eff}})
-S_{0,N_{\rm c}}(T,B)
\nonumber\\
&=S_{0,N_{\rm c}}(T/\lambda,B)
-S_{0,N_{\rm c}}(T,B)
\label{eq:deltaS_fixedN_Beff}\\
&=\int_T^{T/\lambda}
\frac{C_{B,0}(T',B;N_{\rm c})}{T'}\,dT'.
\label{eq:entropy_integral_identity}
\end{align}
Equation~\eqref{eq:entropy_integral_identity} expresses the entropy change
entirely through the Hermitian canonical heat capacity. For positive
$C_{B,0}$ and $|\beta|<1$, one has $T/\lambda\geq T$ and hence
$\Delta S_\beta^{(N_{\rm c})}\geq0$. The sign displayed in
Fig.~\ref{fig:deltaS_fixedN_75} therefore follows directly from the 
scaling relation.

\endgroup

\begingroup

The same scaling determines the temperature change under reversible adiabatic
tuning. For an isentropic process $\beta_i\to\beta_f$ at fixed $B$ and
$N_{\rm c}$,
\begin{equation}
\frac{T_f}{\lambda_f}=\frac{T_i}{\lambda_i},
\label{eq:adiabatic_scaling}
\end{equation}
and therefore
\begin{equation}
T_f=T_i\frac{\lambda_f}{\lambda_i},
\qquad
\Delta T_{\mathrm{ad}}
=T_i\left(\frac{\lambda_f}{\lambda_i}-1\right).
\label{eq:adiabatic_temperature}
\end{equation}
Its differential form is
\begin{equation}
\left(\frac{\partial T}{\partial\beta}\right)_{S,N_{\rm c},B}
=-\frac{\beta T}{1-\beta^2}.
\label{eq:differential_caloric_coefficient}
\end{equation}
Within the regular real-spectrum interval, increasing $|\beta|$ lowers the
temperature along finite-temperature isentropes with monotonic entropy.

\endgroup

\subsection{Fixed chemical potential}
\label{subsec:caloric_fixedmu}

\begin{figure}[t]
\centering
\includegraphics[width=\linewidth]{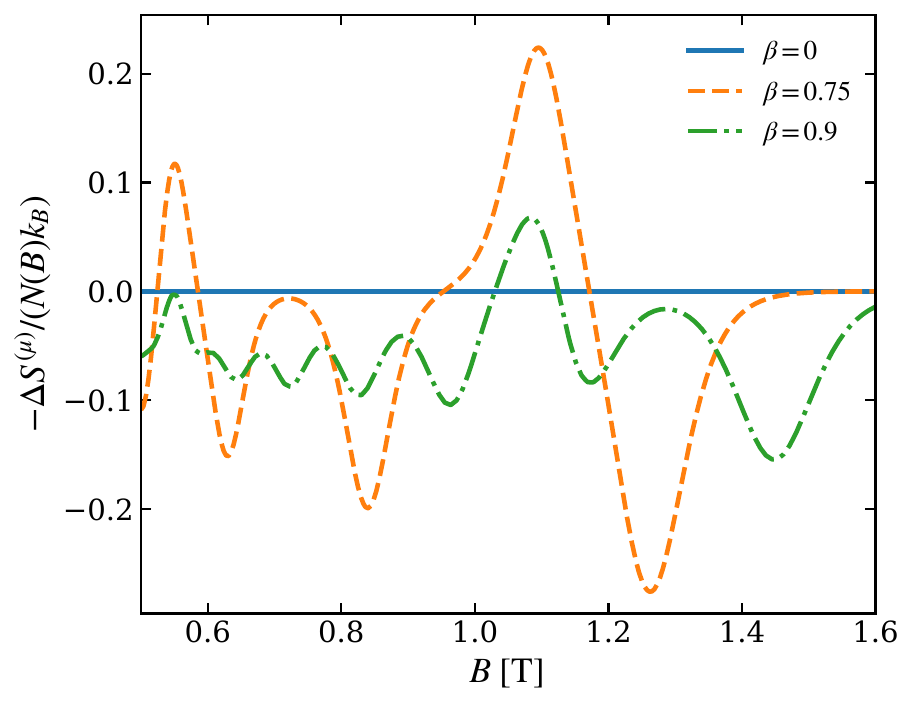}
\caption{Normalized fixed-$\mu$ entropy change at
$T=5~\mathrm K$ and $\mu=38~\mathrm{meV}$.  The sign reversals arise from the
difference between the same Hermitian entropy curve evaluated at $B$ and
$B_{\mathrm{eff}}$.}
\label{fig:deltaS_mufixed}
\end{figure}

\begingroup

We next consider the grand-canonical protocol, in which the reservoir fixes
$\mu$ while $\beta$ is varied. The normalized isothermal entropy change is
\begin{equation}
-\Delta s_\beta^{(\mu)}(B)
=-\frac{S(T,\mu;B,\beta)-S(T,\mu;B,0)}
{N_\beta(B)k_{\mathrm B}},
\label{eq:deltaS_fixedmu}
\end{equation}
where $N_\beta(B)=N_{0,+}(T,\mu;B,\beta)$. The effective-field mapping gives
\begin{align}
\Delta S_\beta^{(\mu)}(T,\mu;B)
&=S_0(T,\mu;B_{\mathrm{eff}})
-S_0(T,\mu;B),
\label{eq:deltaS_fixedmu_Beff}\\
N_\beta(T,\mu;B)
&=N_0(T,\mu;B_{\mathrm{eff}}).
\label{eq:N_fixedmu_Beff_main}
\end{align}
The entropy change therefore compares the Hermitian response at the applied
field $B$ and at the compressed field $B_{\mathrm{eff}}$. As these two fields
sample different positions within the Landau-level oscillations, the entropy
change alternates in sign when successive levels pass through the reservoir
energy.

Figure~\ref{fig:deltaS_mufixed} displays this behavior. Its extrema occur near
the same Landau-level crossings that organize the heat-capacity oscillations
in Fig.~\ref{fig:Cmu_fixedmu}, with the two figures sampling different field
windows and Landau-level indices. For an isolated level, the fixed-$\mu$ heat
capacity vanishes at the crossing and reaches its largest values on the
thermally broadened flanks.

Particle exchange with the reservoir also contributes to the energy balance.
According to Eq.~\eqref{eq:first_law_beta}, the corresponding change in
internal energy combines the reversible heat $T\Delta S$, the chemical term
$\mu\Delta N$, and the work associated with tuning $\beta$.

\endgroup

\section{Discussion and conclusions}
\label{sec:discussion_conclusions}

\begingroup

We have derived scaling relations for the equilibrium thermodynamics of a
real-spectrum NH Dirac Hamiltonian in a magnetic field. The similarity
transformation
in Eq.~\eqref{eq:similarity_H} maps the deformed Hamiltonian
at fixed applied field onto a Hermitian Dirac Hamiltonian with velocity
$\lambda v$. For observables governed by the Landau-level spectrum, the same spectral compression
is represented by the effective field
$B_{\mathrm{eff}}=\lambda^2B$. This structure determines the chemical
potential, entropy, heat capacities, and orbital magnetic response from the
corresponding Hermitian thermodynamic functions.

The thermodynamic signatures depend on the imposed constraint. At fixed PFF, the chemical potential adjusts self-consistently to the compressed Landau spectrum, and the canonical responses follow from the corresponding Hermitian functions under a rescaling of temperature or field. At fixed chemical potential, Landau levels successively cross the reservoir energy, producing oscillatory caloric and magnetic responses. Their field positions encode the compression factor, while their line shapes remain those of the Hermitian representative.

The field dependence of the orbital degeneracy is central to the magnetic
response. The total orbital moment of the projected electron sector obeys an
effective-field scaling, whereas the moment per projected electron carries an
additional factor $\lambda^2$. Quasistatic tuning of $\beta$ also produces a
caloric response governed by the same spectral compression. At fixed projected
filling factor, the entropy change is determined by the canonical heat
capacity, and an isentropic variation of $\beta$ yields the temperature
coefficient in Eq.~\eqref{eq:differential_caloric_coefficient}.
The magnetic scaling relations refer to the projected $(0,+)$
electron-sector contribution relative to the filled valence band; the complete
orbital magnetization additionally requires the regularized valence-band
background.

The equilibrium formulation assumes a bath or reservoir that realizes the
Gibbs ensemble associated with the quasi-Hermitian Hamiltonian. The
Landau-level structure is resolved when
\begin{equation}
k_{\mathrm B}T,\ \Gamma \ll \Delta_n(\beta),
\label{eq:LL_resolution_condition}
\end{equation}
where $\Gamma$ is the single-particle linewidth and $\Delta_n(\beta)$ is given
in Eq.~\eqref{eq:LL_spacing}. Since the relativistic level spacing decreases
with the Landau-level index, broadening first suppresses the high-index
oscillations. As $|\beta|$ increases, the compressed level
spacing eventually becomes comparable to $k_{\mathrm B}T$ or $\Gamma$, so
finite temperature and linewidth limit the experimentally resolved scaling
before the singular threshold $|\beta|=1$ is reached.

The scaling relations provide a reference for isolating thermodynamic effects
associated with additional energy scales. Zeeman splitting, substrate-induced
masses, second-neighbor hopping $t_2$, energy-dependent self-energies,
linewidth variations, and interactions generally introduce distinct
dependencies on $B$ and $\beta$, producing departures from
Eq.~\eqref{eq:main_scaling_relations}. As $|\beta|\to1$, the scaling framework
reaches the boundary of the real-spectrum sector: the metric and similarity
transformation become singular, the Landau spectrum collapses, and the
spectrum becomes increasingly sensitive to
perturbations~\cite{TrefethenEmbree2005,jezequel2026when}.
\endgroup

\begin{acknowledgments} 
This work is supported by Fondecyt (Chile) Grants No.~1230933 (V.J.), No.~1240582 (P.V.), and No.~1250173 (F.J.P.). F. J.P. acknowledges support from CNPq 446761/2025-7. J.P.E. acknowledges support from Agencia Nacional de Investigaci\'on y Desarrollo (ANID) through the Doctorado Nacional Grant No.~2024-21240412. B.C. and J.P.E. acknowledge support from PUCV. B.C. acknowledges support from the Direcci\'on de Postgrado of UTFSM. B.C. further acknowledges support from the Programa de Incentivo a la Iniciaci\'on Cient\'ifica (PIIC) No.~006/2026 and ANID Becas/Doctorado Nacional Grant No.~21250015. P.V. also acknowledges support from CEDENNA under Grant CIA No.~250002. 
\end{acknowledgments}


\appendix

\section{Zero-field lattice benchmark}
\label{app:pristine}

\subsection{Tight-binding spectrum and density of states}
\label{sec:pristine_DOS}

\begingroup

We consider the nearest-neighbor tight-binding Hamiltonian of graphene,
\begin{equation}
H_{\mathrm{TB},0}(\mathbf{k})
=
-t
\begin{pmatrix}
0 & f(\mathbf{k})\\
f^{*}(\mathbf{k}) & 0
\end{pmatrix},
\qquad
f(\mathbf{k})
=
\sum_{j=1}^{3}e^{i\mathbf{k}\cdot\bm{\delta}_{j}},
\label{eq:TB_Hermitian_t2zero}
\end{equation}
where $t\simeq3~\mathrm{eV}$, $a=1.42~\text{\AA}$, and
\begin{equation}
\bm{\delta}_1=a(1,0),
\qquad
\bm{\delta}_{2,3}
=a\left(-\frac12,\pm\frac{\sqrt3}{2}\right).
\label{eq:TB_nearest_neighbor_vectors}
\end{equation}
For $t_2=0$, the model has sublattice symmetry,
\begin{equation}
\{\sigma_z,H_{\mathrm{TB},0}(\mathbf{k})\}=0.
\label{eq:TB_chiral_symmetry}
\end{equation}

The NH deformation is introduced as
\begin{equation}
H_{\mathrm{TB},\beta}(\mathbf{k})
=
(\bm{1}+\beta\sigma_z)H_{\mathrm{TB},0}(\mathbf{k}).
\label{eq:TB_NH_Hamiltonian}
\end{equation}
Using Eq.~\eqref{eq:TB_chiral_symmetry}, one finds
\begin{equation}
H_{\mathrm{TB},\beta}^{2}(\mathbf{k})
=(1-\beta^2)H_{\mathrm{TB},0}^{2}(\mathbf{k}),
\label{eq:TB_NH_squared}
\end{equation}
and hence, for $|\beta|<1$,
\begin{align}
E_{\pm,\beta}(\mathbf{k})
&=
\lambda(\beta)E_{\pm,0}(\mathbf{k}),
\label{eq:TB_NH_bands}\\
\lambda(\beta)&=\sqrt{1-\beta^2}.
\label{eq:TB_NH_bands-1}
\end{align}
The deformation therefore compresses the entire nearest-neighbor spectrum by
the common factor $\lambda(\beta)$. Near the Dirac points, this corresponds to
\begin{equation}
v_F(\beta)=\lambda(\beta)v_F(0).
\label{eq:TB_velocity_scaling}
\end{equation}

\endgroup

\begin{figure}[t]
\centering
\includegraphics[width=\linewidth]{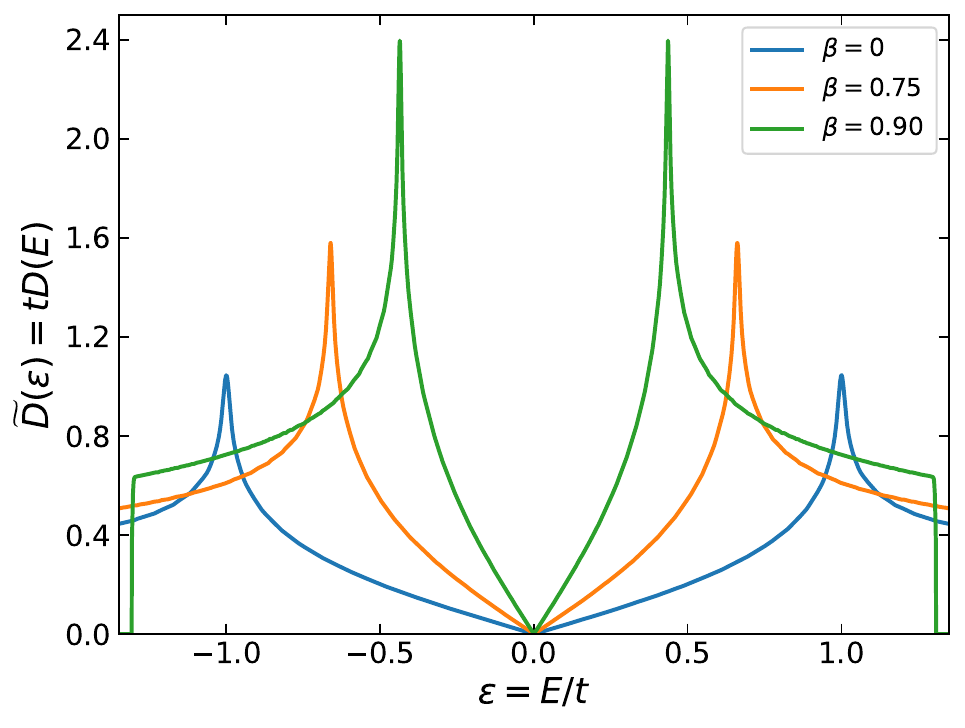}
\caption{Dimensionless density of states
$\widetilde D(\varepsilon)=tD(E)$, with $\varepsilon=E/t$, for representative
values of $\beta$. It is normalized as
$\int d\varepsilon\,\widetilde D(\varepsilon)=2$,
corresponding to two single-particle states per primitive
cell per spin.
For $t_2=0$, the Dirac-point and van Hove structures retain
their Hermitian form while their energies are compressed by
$\lambda(\beta)$.}
\label{fig:DOS_TB_graphene}
\end{figure}

Conservation of the total spectral weight gives the corresponding DOS scaling,
\begin{equation}
D_{\beta}(E)
=
\frac{1}{\lambda(\beta)}
D_0\!\left(
\frac{E}{\lambda(\beta)}
\right).
\label{eq:TB_DOS_scaling}
\end{equation}
Thus, the Dirac and van Hove structures retain their Hermitian form while
shifting to lower energies, as shown in Fig.~\ref{fig:DOS_TB_graphene}.

The full-Brillouin-zone DOS includes both valleys and is
displayed per spin, whereas the continuum expression below includes the spin
and valley multiplicities explicitly through $g=4$.

\subsection{Heat capacity and spectral compression}
\label{sec:cv_pristine}

We next use the full-band DOS to evaluate the zero-field thermal response at
$t_2=0$ and $\mu=0$. Particle--hole symmetry pins the chemical potential at
zero.

\begingroup

In the low-temperature Dirac regime, the positive-energy DOS per unit area is
\begin{equation}
\frac{D_{\beta,+}(E)}{A}
=
\frac{gE}{2\pi\hbar^2v^2\lambda^2},
\qquad E>0.
\label{eq:zero_field_Dirac_DOS}
\end{equation}
Including the electron and hole contributions at charge neutrality gives
\begin{equation}
\frac{C_\beta}{A}
=
\frac{9g\zeta(3)}{2\pi}
\frac{k_{\mathrm B}^3T^2}{\hbar^2v^2\lambda^2}
=
\frac{1}{\lambda^2}\frac{C_0}{A}.
\label{eq:zero_field_lowT_heat_capacity}
\end{equation}
Spectral compression therefore enhances the low-temperature electronic
heat capacity by the factor $\lambda^{-2}$.

\endgroup

For the full tight-binding spectrum, the heat capacity is evaluated from
\begin{equation}
C(T)
=
\left(
\frac{\partial U}{\partial T}
\right)_{\mu=0},
\label{eq:TB_specific_heat}
\end{equation}
and reported as
\begin{equation}
\frac{C(T)}{Nk_{\mathrm B}}.
\label{eq:TB_specific_heat_normalized}
\end{equation}
Here
$N=\int dE\,D(E)f(E,0,T)$ is the total occupation associated with the
normalized full-band DOS.

\begin{figure}[t]
\centering
\includegraphics[width=\linewidth]{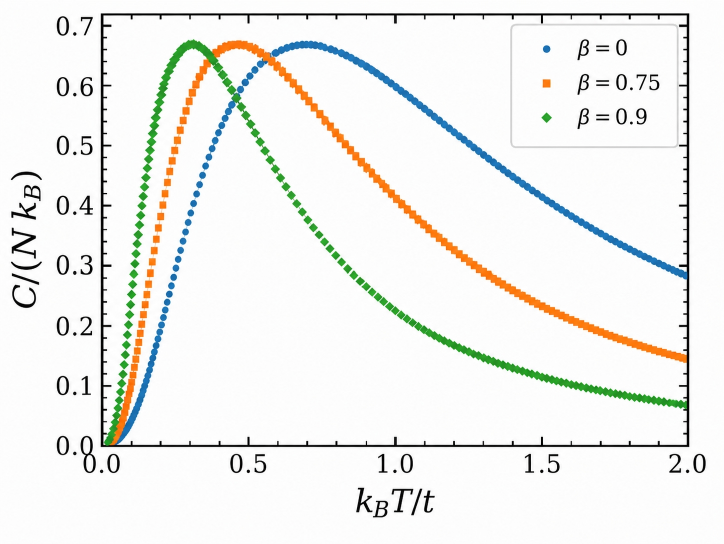}
\caption{Dimensionless electronic heat capacity of
pristine graphene at $B=0$, $C/(Nk_{\mathrm B})$, as a function of
$k_{\mathrm B}T/t$. All curves use the particle--hole-symmetric model with
second-neighbor hopping $t_2=0$. Increasing $\beta$ shifts the full-band maximum
toward lower temperatures according to the spectral compression in
Eq.~\eqref{eq:TB_NH_bands}.}
\label{fig:CvsTbeta_pristine}
\end{figure}

Figure~\ref{fig:CvsTbeta_pristine} shows that the full-band maximum shifts
toward lower temperatures as the spectrum is compressed. The zero-field result
provides a lattice-scale reference for the magnetic-field thermal and orbital responses discussed in the main text.
\section{Scaling relations under uniform spectral compression}
\label{app:scaling}

\begingroup

This appendix derives the relations summarized in
Eq.~\eqref{eq:main_scaling_relations} and applies them to the two ensemble
constraints and to the electron-sector orbital magnetic response.
\endgroup

The deformation considered throughout this work produces the  spectral rescaling
\begin{equation}
E_n(\beta)
=
\lambda E_n(0),
\qquad
\lambda
=
\sqrt{1-\beta^2}.
\label{eq:appA_spectral_compression}
\end{equation}

At fixed magnetic field, the corresponding reduced DOS satisfies
\begin{equation}
D_\beta(E;B)
=
\frac{1}{\lambda}
D_0\!\left(
\frac{E}{\lambda};B
\right).
\label{eq:appA_DOS_scaling}
\end{equation}
For a common Landau-level cutoff, or for a finite-band
lattice regularization, the change of variables preserves the integrated
spectral weight,
\begin{equation}
\int dE\,D_\beta(E;B)
=
\int dE\,D_0(E;B).
\end{equation}
Equivalently, the mapping preserves the number, ordering, and
indexing of the retained Landau levels.

\subsection{Effective magnetic field}
\begingroup

For the relativistic Landau levels, the energy rescaling also has a representation in terms of an 
effective magnetic field. With \(B_{\mathrm{eff}}=\lambda^2B\) from
Eq.~\eqref{eq:Beff},
\begin{equation}
D_\beta(E;B)
=
D_0(E;B_{\mathrm{eff}})
=
\frac{1}{\lambda}D_0\!\left(\frac{E}{\lambda};B\right).
\label{eq:app_effective_field_DOS}
\end{equation}
It follows directly that, at fixed \(T\) and \(\mu\),
\begin{align}
N_\beta(T,\mu;B)
&=
N_0(T,\mu;B_{\mathrm{eff}}),
\label{eq:app_N_Beff}
\\
U_\beta(T,\mu;B)
&=
U_0(T,\mu;B_{\mathrm{eff}}),
\label{eq:app_U_Beff}
\\
S_\beta(T,\mu;B)
&=
S_0(T,\mu;B_{\mathrm{eff}}),
\label{eq:app_S_Beff}
\\
\Omega_\beta(T,\mu;B)
&=
\Omega_0(T,\mu;B_{\mathrm{eff}}),
\label{eq:app_Omega_Beff}
\\
\mathcal C_{\mu,\beta}(T,\mu;B)
&=
\mathcal C_{\mu,0}(T,\mu;B_{\mathrm{eff}}),
\label{eq:app_Cmu_Beff}
\\
c_{\mu,\beta}(T,\mu;B)
&=
c_{\mu,0}(T,\mu;B_{\mathrm{eff}}).
\label{eq:app_cmu_Beff}
\end{align}
Thus the fixed-\(\mu\) curves are not independent deformations of the
Hermitian response: they are the same reduced curves evaluated at
\(B_{\mathrm{eff}}\). For
\(\Phi\in\{N,U,S,\Omega,\mathcal C_\mu\}\), the corresponding projected-sector physical
extensive quantity obeys
\begin{equation}
\Phi_\beta^{\mathrm{phys}}(T,\mu;B)
=
\lambda^{-2}
\Phi_0^{\mathrm{phys}}(T,\mu;B_{\mathrm{eff}}),
\label{eq:app_physical_Beff}
\end{equation}
because
\(\mathcal D_{B_{\mathrm{eff}}}=\lambda^2\mathcal D_B\). This prefactor is
precisely why \(B_{\mathrm{eff}}\) should be understood as a variable encoding  the spectrum
collapse rather than as a replacement for the applied field.
\endgroup

\subsection{Fixed chemical potential}

Changing variables from \(E\) to \(x=E/\lambda\) gives
\begin{align}
N_\beta(T,\mu;B)
&=
N_0\!\left(
\frac{T}{\lambda},
\frac{\mu}{\lambda};B
\right),
\label{eq:appA_N_scaling}
\\
U_\beta(T,\mu;B)
&=
\lambda
U_0\!\left(
\frac{T}{\lambda},
\frac{\mu}{\lambda};B
\right),
\label{eq:appA_U_scaling}
\\
S_\beta(T,\mu;B)
&=
S_0\!\left(
\frac{T}{\lambda},
\frac{\mu}{\lambda};B
\right),
\label{eq:appA_S_scaling}
\\
\Omega_\beta(T,\mu;B)
&=
\lambda
\Omega_0\!\left(
\frac{T}{\lambda},
\frac{\mu}{\lambda};B
\right).
\label{eq:appA_Omega_scaling}
\end{align}
The grand-canonical heat capacity
\begin{equation}
\mathcal C_{\mu,\beta}
=
T
\left(
\frac{\partial S_\beta}{\partial T}
\right)_{\mu,B}
\end{equation}
therefore obeys
\begin{equation}
\mathcal C_{\mu,\beta}(T,\mu;B)
=
\mathcal C_{\mu,0}\!\left(
\frac{T}{\lambda},
\frac{\mu}{\lambda};B
\right).
\label{eq:appA_Cmu_scaling}
\end{equation}

Since the occupation transforms under the same reparametrization, the heat
capacity per particle satisfies
\begin{equation}
c_{\mu,\beta}(T,\mu;B)
=
c_{\mu,0}\!\left(
\frac{T}{\lambda},
\frac{\mu}{\lambda};B
\right).
\label{eq:appA_cmu_scaling}
\end{equation}

\subsection{Fixed projected filling factor}

Let \(\mu_\beta(T,B;N_{\rm c})\) be the solution of
\begin{equation}
N_\beta(T,\mu_\beta;B)
=
N_{\rm c}.
\end{equation}

For \(T>0\) and a prescribed occupation inside the
finite spectral range, \(N_\beta\) is strictly increasing with \(\mu\), so the
solution is unique; the \(T\to0\) result is understood as the corresponding
limit. Using Eq.~\eqref{eq:appA_N_scaling}, this gives
\begin{equation}
\mu_\beta(T,B;N_{\rm c})
=
\lambda
\mu_0\!\left(
\frac{T}{\lambda},
B;N_{\rm c}
\right).
\label{eq:appA_mu_fixedN_scaling}
\end{equation}

Substitution along the self-consistent trajectory yields
\begin{align}
U_{\beta,N_{\rm c}}(T,B)
&=
\lambda
U_{0,N_{\rm c}}\!\left(
\frac{T}{\lambda},B
\right),
\label{eq:appA_U_fixedN_scaling}
\\
S_{\beta,N_{\rm c}}(T,B)
&=
S_{0,N_{\rm c}}\!\left(
\frac{T}{\lambda},B
\right),
\label{eq:appA_S_fixedN_scaling}
\\
C_{B,\beta}(T,B;N_{\rm c})
&=
C_{B,0}\!\left(
\frac{T}{\lambda},B;N_{\rm c}
\right).
\label{eq:appA_CB_fixedN_scaling}
\end{align}

Thus, the self-consistent chemical potential does not break
the spectral scaling. Instead, it follows the same rescaling and fixes the
corresponding canonical temperature reparametrization.

\begingroup

The same canonical identities take the equivalent effective-field form
\begin{align}
\mu_\beta(T,B;N_{\rm c})
&=
\mu_0(T,B_{\mathrm{eff}};N_{\rm c}),
\label{eq:app_mu_fixedN_Beff}
\\
U_{\beta,N_{\rm c}}(T,B)
&=
U_{0,N_{\rm c}}(T,B_{\mathrm{eff}}),
\label{eq:app_U_fixedN_Beff}
\\
S_{\beta,N_{\rm c}}(T,B)
&=
S_{0,N_{\rm c}}(T,B_{\mathrm{eff}}),
\label{eq:app_S_fixedN_Beff}
\\
F_{\beta,N_{\rm c}}(T,B)
&=
F_{0,N_{\rm c}}(T,B_{\mathrm{eff}}),
\label{eq:app_F_fixedN_Beff}
\\
C_{B,\beta}(T,B;N_{\rm c})
&=
C_{B,0}(T,B_{\mathrm{eff}};N_{\rm c}).
\label{eq:app_CB_fixedN_Beff}
\end{align}
These identities hold at fixed PFF. During a field sweep,
\(N_{\rm c}\) remains constant, while the physical particle number varies as
\(N_{0,+}^{\mathrm{phys}}=\mathcal D_BN_{\rm c}\).
\endgroup

\subsection{Magnetic-field observables}

\begingroup

From Eq.~\eqref{eq:app_Omega_Beff} and
\(\partial B_{\mathrm{eff}}/\partial B=\lambda^2\), the reduced spectral orbital response satisfies
\begin{equation}
M_{\mathrm{sp},\beta}(T,\mu;B)
=
\lambda^2
M_{\mathrm{sp},0}(T,\mu;B_{\mathrm{eff}}).
\label{eq:app_Msp_Beff}
\end{equation}
The extensive projected grand potential instead obeys
\begin{equation}
\Omega_\beta^{\mathrm{phys}}(T,\mu;B)
=
\lambda^{-2}
\Omega_0^{\mathrm{phys}}(T,\mu;B_{\mathrm{eff}}),
\label{eq:app_Omega_physical_Beff}
\end{equation}
because
\(\mathcal D_{B_{\mathrm{eff}}}=\lambda^2\mathcal D_B\).
Differentiation with respect to the physical field therefore gives
\begin{equation}
M_{0,+,\beta}^{\mathrm{phys}}(T,\mu;B)
=
M_{0,+,0}^{\mathrm{phys}}(T,\mu;B_{\mathrm{eff}}).
\label{eq:app_Mphysical_Beff}
\end{equation}
Likewise,
\begin{equation}
N_{0,+,\beta}^{\mathrm{phys}}(T,\mu;B)
=
\lambda^{-2}
N_{0,+,0}^{\mathrm{phys}}(T,\mu;B_{\mathrm{eff}}),
\label{eq:app_Nphysical_Beff}
\end{equation}
so the orbital magnetic moment per particle obeys
\begin{equation}
m_{0,+,\beta}^{\mathrm{phys}}(T,\mu;B)
=
\lambda^2
m_{0,+,0}^{\mathrm{phys}}(T,\mu;B_{\mathrm{eff}}).
\label{eq:app_mphysical_Beff}
\end{equation}
Hence the total electron-sector orbital magnetic moment collapses under the horizontal
rescaling \(B\mapsto B_{\mathrm{eff}}\), while the per-particle response
collapses when \(m_{0,+,\beta}^{\mathrm{phys}}/\lambda^2\) is plotted against
\(B_{\mathrm{eff}}\). These identities hold for the projected \((0,+)\)
sector; they do not supply the omitted regularized valence contribution.
\endgroup

\bibliography{nonhermitian_graphene_equilibrium_thermo}

@article{BenderBoettcher1998PRL,
  author  = {Bender, Carl M. and Boettcher, Stefan},
  title   = {Real Spectra in Non-Hermitian Hamiltonians Having {PT} Symmetry},
  journal = {Physical Review Letters},
  volume  = {80},
  pages   = {5243--5246},
  year    = {1998},
  doi     = {10.1103/PhysRevLett.80.5243}
}

@article{ElGanainy2018NatPhys,
  author  = {El-Ganainy, Ramy and Makris, Konstantinos G. and Khajavikhan, Mercedeh and Musslimani, Ziad H. and Rotter, Stefan and Christodoulides, Demetrios N.},
  title   = {Non-Hermitian physics and {PT} symmetry},
  journal = {Nature Physics},
  volume  = {14},
  pages   = {11--19},
  year    = {2018},
  doi     = {10.1038/nphys4323}
}

@article{GoriniKossakowskiSudarshan1976JMP,
  author  = {Gorini, Vittorio and Kossakowski, Andrzej and
             Sudarshan, E. C. G.},
  title   = {Completely Positive Dynamical Semigroups of
             {$N$}-Level Systems},
  journal = {J. Math. Phys.},
  volume  = {17},
  number  = {5},
  pages   = {821--825},
  year    = {1976},
  doi     = {10.1063/1.522979}
}

@article{Scholtz1992AnnPhys,
  author  = {Scholtz, F. G. and Geyer, H. B. and Hahne, F. J. W.},
  title   = {Quasi-Hermitian Operators in Quantum Mechanics and the
             Variational Principle},
  journal = {Annals of Physics},
  volume  = {213},
  number  = {1},
  pages   = {74--101},
  year    = {1992},
  doi     = {10.1016/0003-4916(92)90284-S}
}

@article{Lindblad1976CMP,
  author  = {Lindblad, G{\"o}ran},
  title   = {On the Generators of Quantum Dynamical Semigroups},
  journal = {Commun. Math. Phys.},
  volume  = {48},
  number  = {2},
  pages   = {119--130},
  year    = {1976},
  doi     = {10.1007/BF01608499}
}

@article{Ashida2020AdvPhys,
  author  = {Ashida, Yuto and Gong, Zongping and Ueda, Masahito},
  title   = {Non-Hermitian physics},
  journal = {Advances in Physics},
  volume  = {69},
  number  = {3},
  pages   = {249--435},
  year    = {2020},
  doi     = {10.1080/00018732.2021.1876991}
}

@article{Bergholtz2021RMP,
  author  = {Bergholtz, Emil J. and Budich, Jan Carl and Kunst, Flore K.},
  title   = {Exceptional topology of non-Hermitian systems},
  journal = {Reviews of Modern Physics},
  volume  = {93},
  pages   = {015005},
  year    = {2021},
  doi     = {10.1103/RevModPhys.93.015005}
}

@article{Kawabata2019PRX,
  author  = {Kawabata, Kohei and Shiozaki, Ken and Ueda, Masahito and Sato, Masatoshi},
  title   = {Symmetry and topology in non-Hermitian physics},
  journal = {Physical Review X},
  volume  = {9},
  pages   = {041015},
  year    = {2019},
  doi     = {10.1103/PhysRevX.9.041015}
}

@article{Torres2019JPhysMater,
  author  = {Foa Torres, Luis E. F.},
  title   = {Perspective on topological states of non-Hermitian lattices},
  journal = {Journal of Physics: Materials},
  volume  = {3},
  pages   = {014002},
  year    = {2019},
  doi     = {10.1088/2515-7639/ab4092}
}

@article{Gong2018PRX,
  author  = {Gong, Zongping and Ashida, Yuto and Kawabata, Kohei and Takasan, Kazuaki and Higashikawa, Sho and Ueda, Masahito},
  title   = {Topological Phases of Non-Hermitian Systems},
  journal = {Physical Review X},
  volume  = {8},
  pages   = {031079},
  year    = {2018},
  doi     = {10.1103/PhysRevX.8.031079}
}

@article{ShenZhenFu2018PRL,
  author  = {Shen, Huitao and Zhen, Bo and Fu, Liang},
  title   = {Topological Band Theory for Non-Hermitian Hamiltonians},
  journal = {Physical Review Letters},
  volume  = {120},
  pages   = {146402},
  year    = {2018},
  doi     = {10.1103/PhysRevLett.120.146402}
}

@article{Kunst2018PRL,
  author  = {Kunst, Flore K. and Edvardsson, Elisabet and Budich, Jan Carl and Bergholtz, Emil J.},
  title   = {Biorthogonal Bulk-Boundary Correspondence in Non-Hermitian Systems},
  journal = {Physical Review Letters},
  volume  = {121},
  pages   = {026808},
  year    = {2018},
  doi     = {10.1103/PhysRevLett.121.026808}
}

@article{Mostafazadeh2002PsHI,
    author = {Mostafazadeh, Ali},
    title = {Pseudo-Hermiticity versus PT symmetry: The necessary condition for the reality of the spectrum of a non-Hermitian Hamiltonian},
    journal = {Journal of Mathematical Physics},
    volume = {43},
    number = {1},
    pages = {205-214},
    year = {2002},
    month = {01},
    issn = {0022-2488},
    doi = {10.1063/1.1418246},
    url = {https://doi.org/10.1063/1.1418246},
}

@article{Mostafazadeh2002PsHII,
  author  = {Mostafazadeh, Ali},
  title   = {Pseudo-Hermiticity versus {PT} symmetry. {II}. A complete characterization of non-Hermitian Hamiltonians with a real spectrum},
  journal = {Journal of Mathematical Physics},
  volume  = {43},
  pages   = {2814--2816},
  year    = {2002},
  doi     = {10.1063/1.1461427}
}

@article{Mostafazadeh2010IJGMMP,
  author  = {Mostafazadeh, Ali},
  title   = {Pseudo-Hermitian Representation of Quantum Mechanics},
  journal = {International Journal of Geometric Methods in Modern Physics},
  volume  = {7},
  number  = {7},
  pages   = {1191--1306},
  year    = {2010},
  doi     = {10.1142/S0219887810004816}
}

@article{GardasDeffner2016SciRep,
  author  = {Gardas, Bart{\l}omiej and Deffner, Sebastian and Saxena, Avadh},
  title   = {Non-Hermitian Quantum Thermodynamics},
  journal = {Scientific Reports},
  volume  = {6},
  pages   = {23408},
  year    = {2016},
  doi     = {10.1038/srep23408}
}

@article{Bebiano2020JMP,
  author  = {Bebiano, Natalia and Providencia, Jo{\~a}o da and Providencia, Jo{\~a}o P. da},
  title   = {Toward non-Hermitian quantum statistical thermodynamics},
  journal = {Journal of Mathematical Physics},
  volume  = {61},
  pages   = {022102},
  year    = {2020},
  doi     = {10.1063/1.5122182}
}

@article{CastroNeto2009RMP,
  author  = {Castro Neto, Antonio H. and Guinea, Francisco and Peres, Nuno M. R. and Novoselov, Kostya S. and Geim, Andre K.},
  title   = {The electronic properties of graphene},
  journal = {Reviews of Modern Physics},
  volume  = {81},
  pages   = {109--162},
  year    = {2009},
  doi     = {10.1103/RevModPhys.81.109}
}

@article{Abergel2010AdvPhys,
  author  = {Abergel, D. S. L. and Apalkov, V. and Berashevich, J. and Ziegler, K. and Chakraborty, Tapash},
  title   = {Properties of graphene: a theoretical perspective},
  journal = {Advances in Physics},
  volume  = {59},
  number  = {4},
  pages   = {261--482},
  year    = {2010},
  doi     = {10.1080/00018732.2010.487978}
}

@article{Novoselov2005Nature,
  author  = {Novoselov, K. S. and Geim, A. K. and Morozov, S. V. and Jiang, D. and Katsnelson, M. I. and Grigorieva, I. V. and Dubonos, S. V. and Firsov, A. A.},
  title   = {Two-dimensional gas of massless Dirac fermions in graphene},
  journal = {Nature},
  volume  = {438},
  pages   = {197--200},
  year    = {2005},
  doi     = {10.1038/nature04233}
}

@article{Zhang2005Nature,
  author  = {Zhang, Yuanbo and Tan, Yan-Wen and Stormer, Horst L. and Kim, Philip},
  title   = {Experimental observation of the quantum Hall effect and {Berry's} phase in graphene},
  journal = {Nature},
  volume  = {438},
  pages   = {201--204},
  year    = {2005},
  doi     = {10.1038/nature04235}
}

@article{Goerbig2011RMP,
  author  = {Goerbig, Mark O.},
  title   = {Electronic properties of graphene in a strong magnetic field},
  journal = {Reviews of Modern Physics},
  volume  = {83},
  pages   = {1193--1243},
  year    = {2011},
  doi     = {10.1103/RevModPhys.83.1193}
}

@article{ZhengAndo2002PRB,
  author  = {Zheng, Yisong and Ando, Tsuneya},
  title   = {Hall conductivity of a two-dimensional graphite system},
  journal = {Physical Review B},
  volume  = {65},
  pages   = {245420},
  year    = {2002},
  doi     = {10.1103/PhysRevB.65.245420}
}

@article{GusyninSharapov2005PRL,
  author  = {Gusynin, V. P. and Sharapov, S. G.},
  title   = {Unconventional Integer Quantum Hall Effect in Graphene},
  journal = {Physical Review Letters},
  volume  = {95},
  pages   = {146801},
  year    = {2005},
  doi     = {10.1103/PhysRevLett.95.146801}
}

@article{GusyninSharapov2006PRB,
  author  = {Gusynin, V. P. and Sharapov, S. G.},
  title   = {Transport of Dirac quasiparticles in graphene: Hall and optical conductivities},
  journal = {Physical Review B},
  volume  = {73},
  pages   = {245411},
  year    = {2006},
  doi     = {10.1103/PhysRevB.73.245411}
}

@article{McClure1956PR,
  author  = {McClure, J. W.},
  title   = {Diamagnetism of Graphite},
  journal = {Physical Review},
  volume  = {104},
  pages   = {666--671},
  year    = {1956},
  doi     = {10.1103/PhysRev.104.666}
}

@article{McClure1960PR,
  author  = {McClure, J. W.},
  title   = {Theory of Diamagnetism of Graphite},
  journal = {Physical Review},
  volume  = {119},
  pages   = {606--613},
  year    = {1960},
  doi     = {10.1103/PhysRev.119.606}
}

@article{KoshinoAndo2007PRB,
  author  = {Koshino, Mikito and Ando, Tsuneya},
  title   = {Orbital diamagnetism in multilayer graphenes: Systematic study with the effective mass approximation},
  journal = {Physical Review B},
  volume  = {76},
  pages   = {085425},
  year    = {2007},
  doi     = {10.1103/PhysRevB.76.085425}
}

@article{BagarelloHatano2016RSPA,
  author  = {Bagarello, Fabio and Hatano, Naomichi},
  title   = {{$\mathcal{PT}$}-symmetric graphene under a magnetic field},
  journal = {Proceedings of the Royal Society A},
  volume  = {472},
  pages   = {20160365},
  year    = {2016},
  doi     = {10.1098/rspa.2016.0365}
}

@article{ZhangFranz2020PRL,
  author  = {Zhang, Xiao-Xiao and Franz, Marcel},
  title   = {Non-Hermitian Exceptional Landau Quantization},
  journal = {Physical Review Letters},
  volume  = {124},
  pages   = {046401},
  year    = {2020},
  doi     = {10.1103/PhysRevLett.124.046401}
}

@article{Regensburger2016,
  author  = {Regensburger, Alois and Bersch, Christian and Miri, Mohammad-Ali and Onishchukov, Grigory and Christodoulides, Demetrios N. and Peschel, Ulf},
  title   = {Parity--time synthetic photonic lattices},
  journal = {Nature},
  volume  = {488},
  pages   = {167--171},
  year    = {2012},
  doi     = {10.1038/nature11298}
}

@article{FongSchwab2012PRX,
  author  = {Fong, Kin Chung and Schwab, K. C.},
  title   = {Ultrasensitive and Wide-Bandwidth Thermal Measurements of Graphene at Low Temperatures},
  journal = {Physical Review X},
  volume  = {2},
  pages   = {031006},
  year    = {2012},
  doi     = {10.1103/PhysRevX.2.031006}
}

@article{Betz2012PRL,
  author  = {Betz, A. C. and Jhang, S. H. and Pallecchi, E. and Ferreira, R. and Fève, G. and Berroir, J.-M. and Plaçais, B.},
  title   = {Supercollision Cooling in Undoped Graphene},
  journal = {Physical Review Letters},
  volume  = {109},
  pages   = {056805},
  year    = {2012},
  doi     = {10.1103/PhysRevLett.109.056805}
}

@book{Callen1985,
  author    = {Callen, Herbert B.},
  title     = {Thermodynamics and an Introduction to Thermostatistics},
  edition   = {2},
  publisher = {Wiley},
  address   = {New York},
  year      = {1985}
}

@book{PathriaBeale2011,
  author    = {Pathria, R. K. and Beale, Paul D.},
  title     = {Statistical Mechanics},
  edition   = {3},
  publisher = {Academic Press},
  address   = {Boston},
  year      = {2011}
}

@book{TrefethenEmbree2005,
  title     = {Spectra and Pseudospectra: The Behavior of Nonnormal Matrices and Operators},
  author    = {Trefethen, Lloyd N. and Embree, Mark},
  publisher = {Princeton University Press},
  address   = {Princeton, NJ},
  year      = {2005}
}

@misc{jezequel2026when,
  title         = {When and why non-{H}ermitian eigenvalues miss eigenstates in topological physics},
  author        = {Jezequel, Lucien and Herviou, Lo{\"i}c and Bardarson, Jens H.},
  year          = {2026},
  eprint        = {2601.05234},
  archivePrefix = {arXiv},
  primaryClass  = {cond-mat.mes-hall},
  doi           = {10.48550/arXiv.2601.05234}
}

@article{MontagOzawa2026NHLL,
  author       = {Montag, Anton and Ozawa, Tomoki},
  title        = {Non-{H}ermitian {L}andau {L}evels},
  journal      = {arXiv preprint},
  eprint       = {2605.23613},
  archivePrefix= {arXiv},
  primaryClass = {cond-mat.mes-hall},
  year         = {2026}
}

@article{CaoKou2023PRResearch,
  author  = {Cao, Kui and Kou, Su-Peng},
  title   = {Statistical mechanics for non-{H}ermitian quantum systems},
  journal = {Phys. Rev. Research},
  volume  = {5},
  pages   = {033196},
  year    = {2023},
  doi     = {10.1103/PhysRevResearch.5.033196}
}

@article{JuricicRoy2024CommunPhys,
  author  = {Juri{\v c}i{\'c}, Vladimir and Roy, Bitan},
  title   = {Yukawa-{L}orentz symmetry in non-{H}ermitian {D}irac materials},
  journal = {Commun. Phys.},
  volume  = {7},
  pages   = {169},
  year    = {2024},
  doi     = {10.1038/s42005-024-01629-2}
}

@article{Roy2025PRD,
  author  = {Roy, Bitan},
  title   = {Zero modes and index theorems for non-{H}ermitian {D}irac fermions},
  journal = {Phys. Rev. D},
  volume  = {112},
  pages   = {125016},
  year    = {2025},
  doi={10.1103/frty-lwz7}
}

@article{EsparzaPenaVargasJuricic2026,
  author       = {Esparza, Juan Pablo and Pe{\~n}a, Francisco J. and Vargas, Patricio and Juri{\v c}i{\'c}, Vladimir},
  title        = {Thermodynamic signatures of non-{H}ermiticity in {D}irac materials via quantum capacitance},
  journal      = {arXiv preprint},
  eprint       = {2604.14150},
  archivePrefix= {arXiv},
  primaryClass = {cond-mat.mes-hall},
  year         = {2026}
}

@article{Savoia2016NH-induced,
  title = {Non-Hermiticity-induced wave confinement and guiding in loss-gain-loss three-layer systems},
  author = {Savoia, Silvio and Castaldi, Giuseppe and Galdi, Vincenzo},
  journal = {Phys. Rev. A},
  volume = {94},
  issue = {4},
  pages = {043838},
  numpages = {10},
  year = {2016},
  month = {Oct},
  publisher = {American Physical Society},
  doi = {10.1103/PhysRevA.94.043838},
  url = {https://link.aps.org/doi/10.1103/PhysRevA.94.043838}
}

@article{Takata2018Photonic,
  title = {Photonic Topological Insulating Phase Induced Solely by Gain and Loss},
  author = {Takata, Kenta and Notomi, Masaya},
  journal = {Phys. Rev. Lett.},
  volume = {121},
  issue = {21},
  pages = {213902},
  numpages = {6},
  year = {2018},
  month = {Nov},
  publisher = {American Physical Society},
  doi = {10.1103/PhysRevLett.121.213902},
  url = {https://link.aps.org/doi/10.1103/PhysRevLett.121.213902}
}

@article{Xue2020NHDirac,
  title = {Non-Hermitian Dirac Cones},
  author = {Xue, Haoran and Wang, Qiang and Zhang, Baile and Chong, Y. D.},
  journal = {Phys. Rev. Lett.},
  volume = {124},
  issue = {23},
  pages = {236403},
  numpages = {6},
  year = {2020},
  month = {Jun},
  publisher = {American Physical Society},
  doi = {10.1103/PhysRevLett.124.236403},
  url = {https://link.aps.org/doi/10.1103/PhysRevLett.124.236403}
}

@article{Peng2024AcousticNHDSMs,
    author = {Peng, Mian and Wu, Chaohua and Cui, Zhenxing and Zhang, Xuewei and Wei, Qiang and Yan, Mou and Chen, Gang},
    title = {Acoustic non-Hermitian Dirac states tuned by flexible designed gain and loss},
    journal = {Applied Physics Letters},
    volume = {125},
    number = {19},
    pages = {193101},
    year = {2024},
    month = {11},
    issn = {0003-6951},
    doi = {10.1063/5.0237506},
    url = {https://doi.org/10.1063/5.0237506},
}

@article{Li2022Gain-Loss,
  title = {Gain-Loss-Induced Hybrid Skin-Topological Effect},
  author = {Li, Yaohua and Liang, Chao and Wang, Chenyang and Lu, Cuicui and Liu, Yong-Chun},
  journal = {Phys. Rev. Lett.},
  volume = {128},
  issue = {22},
  pages = {223903},
  numpages = {7},
  year = {2022},
  month = {Jun},
  publisher = {American Physical Society},
  doi = {10.1103/PhysRevLett.128.223903},
  url = {https://link.aps.org/doi/10.1103/PhysRevLett.128.223903}
}

@article{Jiang2024TunableNHSE,
  title = {Tunable non-Hermitian skin effect via gain and loss},
  author = {Jiang, Wen-Cheng and Wu, Hong and Li, Qing-Xu and Li, Jian and Zhu, Jia-Ji},
  journal = {Phys. Rev. B},
  volume = {110},
  issue = {15},
  pages = {155144},
  numpages = {9},
  year = {2024},
  month = {Oct},
  publisher = {American Physical Society},
  doi = {10.1103/PhysRevB.110.155144},
  url = {https://link.aps.org/doi/10.1103/PhysRevB.110.155144}
}

@article{Cornelius2022Spectral,
  title = {Spectral Filtering Induced by Non-Hermitian Evolution with Balanced Gain and Loss: Enhancing Quantum Chaos},
  author = {Cornelius, Julien and Xu, Zhenyu and Saxena, Avadh and Chenu, Aur\'elia and del Campo, Adolfo},
  journal = {Phys. Rev. Lett.},
  volume = {128},
  issue = {19},
  pages = {190402},
  numpages = {6},
  year = {2022},
  month = {May},
  publisher = {American Physical Society},
  doi = {10.1103/PhysRevLett.128.190402},
  url = {https://link.aps.org/doi/10.1103/PhysRevLett.128.190402}
}

@article{Li2023Loss-induced,
  title = {Loss-induced Floquet non-Hermitian skin effect},
  author = {Li, Yaohua and Lu, Cuicui and Zhang, Shuang and Liu, Yong-Chun},
  journal = {Phys. Rev. B},
  volume = {108},
  issue = {22},
  pages = {L220301},
  numpages = {6},
  year = {2023},
  month = {Dec},
  publisher = {American Physical Society},
  doi = {10.1103/PhysRevB.108.L220301},
  url = {https://link.aps.org/doi/10.1103/PhysRevB.108.L220301}
}

@article{Koutserimpas2018Nonreciprocal,
  title = {Nonreciprocal Gain in Non-Hermitian Time-Floquet Systems},
  author = {Koutserimpas, Theodoros T. and Fleury, Romain},
  journal = {Phys. Rev. Lett.},
  volume = {120},
  issue = {8},
  pages = {087401},
  numpages = {6},
  year = {2018},
  month = {Feb},
  publisher = {American Physical Society},
  doi = {10.1103/PhysRevLett.120.087401},
  url = {https://link.aps.org/doi/10.1103/PhysRevLett.120.087401}
}

@article{Chaduteau2026Lindbladian,
  title = {Lindbladian versus Postselected non-Hermitian Topology},
  author = {Chaduteau, Alexandre and Lee, Derek K. K. and Schindler, Frank},
  journal = {Phys. Rev. Lett.},
  volume = {136},
  issue = {1},
  pages = {016603},
  numpages = {6},
  year = {2026},
  month = {Jan},
  publisher = {American Physical Society},
  doi = {10.1103/ljvt-w6hw},
  url = {https://link.aps.org/doi/10.1103/ljvt-w6hw}
}

@article{Hatano1997Vortex,
  title = {Vortex pinning and non-Hermitian quantum mechanics},
  author = {Hatano, Naomichi and Nelson, David R.},
  journal = {Phys. Rev. B},
  volume = {56},
  issue = {14},
  pages = {8651--8673},
  numpages = {0},
  year = {1997},
  month = {Oct},
  publisher = {American Physical Society},
  doi = {10.1103/PhysRevB.56.8651},
  url = {https://link.aps.org/doi/10.1103/PhysRevB.56.8651}
}

@article{Hatano1998NHDelocalization,
  title = {Non-Hermitian delocalization and eigenfunctions},
  author = {Hatano, Naomichi and Nelson, David R.},
  journal = {Phys. Rev. B},
  volume = {58},
  issue = {13},
  pages = {8384--8390},
  numpages = {0},
  year = {1998},
  month = {Oct},
  publisher = {American Physical Society},
  doi = {10.1103/PhysRevB.58.8384},
  url = {https://link.aps.org/doi/10.1103/PhysRevB.58.8384}
}

@article{Liu2020GainAndLoss,
  title = {Gain- and Loss-Induced Topological Insulating Phase in a Non-Hermitian Electrical Circuit},
  author = {Liu, Shuo and Ma, Shaojie and Yang, Cheng and Zhang, Lei and Gao, Wenlong and Xiang, Yuan Jiang and Cui, Tie Jun and Zhang, Shuang},
  journal = {Phys. Rev. Appl.},
  volume = {13},
  issue = {1},
  pages = {014047},
  numpages = {11},
  year = {2020},
  month = {Jan},
  publisher = {American Physical Society},
  doi = {10.1103/PhysRevApplied.13.014047},
  url = {https://link.aps.org/doi/10.1103/PhysRevApplied.13.014047}
}

@article{Shen2024GainAndLoss,
doi = {10.1088/1361-6463/ad0989},
url = {https://doi.org/10.1088/1361-6463/ad0989},
year = {2023},
month = {nov},
publisher = {IOP Publishing},
volume = {57},
number = {6},
pages = {065102},
author = {Shen, Xizhou and Pan, Keyu and Wang, Xiumei and Jiang, Hengxuan and Zhou, Xingping},
title = {Gain and loss induced higher-order exceptional points in a non-Hermitian electrical circuit},
journal = {Journal of Physics D: Applied Physics}
}

@article{Roccati2022NHPhysics,
author = {Roccati, Federico and Palma, G. Massimo and Ciccarello, Francesco and Bagarello, Fabio},
title = {Non-Hermitian Physics and Master Equations},
journal = {Open Systems \& Information Dynamics},
volume = {29},
number = {01},
pages = {2250004},
year = {2022},
doi = {10.1142/S1230161222500044},
URL = {https://doi.org/10.1142/S1230161222500044}
}

@article{Ezawa2019NHBoundary,
  title = {Non-Hermitian boundary and interface states in nonreciprocal higher-order topological metals and electrical circuits},
  author = {Ezawa, Motohiko},
  journal = {Phys. Rev. B},
  volume = {99},
  issue = {12},
  pages = {121411(R)},
  numpages = {5},
  year = {2019},
  month = {Mar},
  publisher = {American Physical Society},
  doi = {10.1103/PhysRevB.99.121411},
  url = {https://link.aps.org/doi/10.1103/PhysRevB.99.121411}
}

@article{Ghaemi-Dizicheh2023Transport,
  title = {Transport effects in non-Hermitian nonreciprocal systems: General approach},
  author = {Ghaemi-Dizicheh, Hamed},
  journal = {Phys. Rev. B},
  volume = {107},
  issue = {12},
  pages = {125155},
  numpages = {10},
  year = {2023},
  month = {Mar},
  publisher = {American Physical Society},
  doi = {10.1103/PhysRevB.107.125155},
  url = {https://link.aps.org/doi/10.1103/PhysRevB.107.125155}
}

@article{Reisenbauer2024NHDynamics,
  title={Non-Hermitian dynamics and non-reciprocity of optically coupled nanoparticles},
  author={Reisenbauer, Manuel and Rudolph, Henning and Egyed, Livia and Hornberger, Klaus and Zasedatelev, Anton V and Abuzarli, Murad and Stickler, Benjamin A and Deli{\'c}, Uro{\v{s}}},
  journal={Nature Physics},
  volume={20},
  number={10},
  pages={1629--1635},
  year={2024},
  publisher={Nature Publishing Group UK London}
}

@article{Jana2025Harnessing,
  title = {Harnessing nonlinearity to tame wave dynamics in nonreciprocal active systems},
  author = {Jana, Sayan and Many Manda, Bertin and Achilleos, Vassos and Frantzeskakis, Dimitrios J. and Sirota, Lea},
  journal = {Phys. Rev. Appl.},
  volume = {24},
  issue = {4},
  pages = {L041005},
  numpages = {6},
  year = {2025},
  month = {Oct},
  publisher = {American Physical Society},
  doi = {10.1103/zkxs-24kn},
  url = {https://link.aps.org/doi/10.1103/zkxs-24kn}
}

@article{Martinez-Alvarez2018NHRobust,
  title = {Non-Hermitian robust edge states in one dimension: Anomalous localization and eigenspace condensation at exceptional points},
  author = {Martinez Alvarez, V. M. and Barrios Vargas, J. E. and Foa Torres, L. E. F.},
  journal = {Phys. Rev. B},
  volume = {97},
  issue = {12},
  pages = {121401(R)},
  numpages = {6},
  year = {2018},
  month = {Mar},
  publisher = {American Physical Society},
  doi = {10.1103/PhysRevB.97.121401},
  url = {https://link.aps.org/doi/10.1103/PhysRevB.97.121401}
}

@article{Yao2018Edge,
  title = {Edge States and Topological Invariants of Non-Hermitian Systems},
  author = {Yao, Shunyu and Wang, Zhong},
  journal = {Phys. Rev. Lett.},
  volume = {121},
  issue = {8},
  pages = {086803},
  numpages = {8},
  year = {2018},
  month = {Aug},
  publisher = {American Physical Society},
  doi = {10.1103/PhysRevLett.121.086803},
  url = {https://link.aps.org/doi/10.1103/PhysRevLett.121.086803}
}

@article{Okuma2020TopologicalNHSE,
  title = {Topological Origin of Non-Hermitian Skin Effects},
  author = {Okuma, Nobuyuki and Kawabata, Kohei and Shiozaki, Ken and Sato, Masatoshi},
  journal = {Phys. Rev. Lett.},
  volume = {124},
  issue = {8},
  pages = {086801},
  numpages = {7},
  year = {2020},
  month = {Feb},
  publisher = {American Physical Society},
  doi = {10.1103/PhysRevLett.124.086801},
  url = {https://link.aps.org/doi/10.1103/PhysRevLett.124.086801}
}

@article{Song2019NHSE,
  title = {Non-Hermitian Skin Effect and Chiral Damping in Open Quantum Systems},
  author = {Song, Fei and Yao, Shunyu and Wang, Zhong},
  journal = {Phys. Rev. Lett.},
  volume = {123},
  issue = {17},
  pages = {170401},
  numpages = {8},
  year = {2019},
  month = {Oct},
  publisher = {American Physical Society},
  doi = {10.1103/PhysRevLett.123.170401},
  url = {https://link.aps.org/doi/10.1103/PhysRevLett.123.170401}
}

@article{Kawabata2020HONHSE,
  title = {Higher-order non-Hermitian skin effect},
  author = {Kawabata, Kohei and Sato, Masatoshi and Shiozaki, Ken},
  journal = {Phys. Rev. B},
  volume = {102},
  issue = {20},
  pages = {205118},
  numpages = {16},
  year = {2020},
  month = {Nov},
  publisher = {American Physical Society},
  doi = {10.1103/PhysRevB.102.205118},
  url = {https://link.aps.org/doi/10.1103/PhysRevB.102.205118}
}

@article{Zhang2021Observation,
  title={Observation of higher-order non-Hermitian skin effect},
  author={Zhang, Xiujuan and Tian, Yuan and Jiang, Jian-Hua and Lu, Ming-Hui and Chen, Yan-Feng},
  journal={Nature communications},
  volume={12},
  number={1},
  pages={5377},
  year={2021},
  publisher={Nature Publishing Group UK London}
}

@article{Li2022DynamicNHSE,
  title = {Dynamic skin effects in non-Hermitian systems},
  author = {Li, Haoshu and Wan, Shaolong},
  journal = {Phys. Rev. B},
  volume = {106},
  issue = {24},
  pages = {L241112},
  numpages = {6},
  year = {2022},
  month = {Dec},
  publisher = {American Physical Society},
  doi = {10.1103/PhysRevB.106.L241112},
  url = {https://link.aps.org/doi/10.1103/PhysRevB.106.L241112}
}

@article{Liang2022DynamicSignaturesofNHSE,
  title = {Dynamic Signatures of Non-Hermitian Skin Effect and Topology in Ultracold Atoms},
  author = {Liang, Qian and Xie, Dizhou and Dong, Zhaoli and Li, Haowei and Li, Hang and Gadway, Bryce and Yi, Wei and Yan, Bo},
  journal = {Phys. Rev. Lett.},
  volume = {129},
  issue = {7},
  pages = {070401},
  numpages = {6},
  year = {2022},
  month = {Aug},
  publisher = {American Physical Society},
  doi = {10.1103/PhysRevLett.129.070401},
  url = {https://link.aps.org/doi/10.1103/PhysRevLett.129.070401}
}

@article{Zhang2022Universal,
  title={Universal non-Hermitian skin effect in two and higher dimensions},
  author={Zhang, Kai and Yang, Zhesen and Fang, Chen},
  journal={Nature communications},
  volume={13},
  number={1},
  pages={2496},
  year={2022},
  publisher={Nature Publishing Group UK London}
}

@article{Li2024ObservationofDNHSE,
  title={Observation of dynamic non-Hermitian skin effects},
  author={Li, Zhen and Wang, Li-Wei and Wang, Xulong and Lin, Zhi-Kang and Ma, Guancong and Jiang, Jian-Hua},
  journal={Nature Communications},
  volume={15},
  number={1},
  pages={6544},
  year={2024},
  publisher={Nature Publishing Group UK London}
}

@article{Yoshida2024NHMSE,
  title = {Non-Hermitian Mott Skin Effect},
  author = {Yoshida, Tsuneya and Zhang, Song-Bo and Neupert, Titus and Kawakami, Norio},
  journal = {Phys. Rev. Lett.},
  volume = {133},
  issue = {7},
  pages = {076502},
  numpages = {8},
  year = {2024},
  month = {Aug},
  publisher = {American Physical Society},
  doi = {10.1103/PhysRevLett.133.076502},
  url = {https://link.aps.org/doi/10.1103/PhysRevLett.133.076502}
}

@article{Heiss2004EPsofNHoperators,
doi = {10.1088/0305-4470/37/6/034},
url = {https://doi.org/10.1088/0305-4470/37/6/034},
year = {2004},
month = {jan},
publisher = {},
volume = {37},
number = {6},
pages = {2455},
author = {W D Heiss},
title = {Exceptional points of non-Hermitian operators},
journal = {Journal of Physics A: Mathematical and General}
}

@article{Muller2008EPsinOPQS,
doi = {10.1088/1751-8113/41/24/244018},
url = {https://doi.org/10.1088/1751-8113/41/24/244018},
year = {2008},
month = {jun},
publisher = {},
volume = {41},
number = {24},
pages = {244018},
author = {Müller, Markus and Rotter, Ingrid},
title = {Exceptional points in open quantum systems},
journal = {Journal of Physics A: Mathematical and Theoretical}
}

@article{Heiss2012ThephysicsofEPs,
doi = {10.1088/1751-8113/45/44/444016},
url = {https://doi.org/10.1088/1751-8113/45/44/444016},
year = {2012},
month = {oct},
publisher = {IOP Publishing},
volume = {45},
number = {44},
pages = {444016},
author = {Heiss, W D},
title = {The physics of exceptional points},
journal = {Journal of Physics A: Mathematical and Theoretical}
}

@article{Okugawa2019Topological,
  title = {Topological exceptional surfaces in non-Hermitian systems with parity-time and parity-particle-hole symmetries},
  author = {Okugawa, Ryo and Yokoyama, Takehito},
  journal = {Phys. Rev. B},
  volume = {99},
  issue = {4},
  pages = {041202(R)},
  numpages = {6},
  year = {2019},
  month = {Jan},
  publisher = {American Physical Society},
  doi = {10.1103/PhysRevB.99.041202},
  url = {https://link.aps.org/doi/10.1103/PhysRevB.99.041202}
}

@article{Mohammad-Ali2019EPs,
  author  = {Miri, Mohammad-Ali and Al{\`u}, Andrea},
  title   = {Exceptional points in optics and photonics},
  journal = {Science},
  volume  = {363},
  number  = {6422},
  pages   = {eaar7709},
  year    = {2019},
  doi     = {10.1126/science.aar7709}
}

@article{Wu2025Experimental,
  title = {Experimental Observation of Dirac Exceptional Points},
  author = {Wu, Yang and Zhu, Dongfanghao and Wang, Yunhan and Rong, Xing and Du, Jiangfeng},
  journal = {Phys. Rev. Lett.},
  volume = {134},
  issue = {15},
  pages = {153601},
  numpages = {7},
  year = {2025},
  month = {Apr},
  publisher = {American Physical Society},
  doi = {10.1103/PhysRevLett.134.153601},
  url = {https://link.aps.org/doi/10.1103/PhysRevLett.134.153601}
}

@article{Yoshida2026HopfEPs,
	title = {Hopf exceptional points},
	pages = {001},
	author = {Yoshida, Tsuneya and Bergholtz, Emil and Bzdušek, Tomáš},
	journal = {SciPost Phys.},
	volume = {20},
	year = {2026},
	publisher = {SciPost},
	doi = {10.21468/SciPostPhys.20.1.001},
	url = {https://scipost.org/10.21468/SciPostPhys.20.1.001}
}

@article{Li2026exceptional,
  title     = {Exceptional deficiency of non-{H}ermitian systems},
  author    = {Li, Zhen and Cai, Rundong and Wang, Xulong and Shimomura, Kenji and Lu, Congwei and Yang, Zhesen and Sato, Masatoshi and Ma, Guancong},
  journal   = {Nature Physics},
  volume    = {22},
  number    = {6},
  pages     = {962--970},
  year      = {2026},
  month     = {Jun},
  doi       = {10.1038/s41567-026-03259-7}
}

@article{Kawabata2019Topological,
  title={Topological unification of time-reversal and particle-hole symmetries in non-Hermitian physics},
  author={Kawabata, Kohei and Higashikawa, Sho and Gong, Zongping and Ashida, Yuto and Ueda, Masahito},
  journal={Nature communications},
  volume={10},
  number={1},
  pages={297},
  year={2019},
  publisher={Nature Publishing Group UK London}
}

@article{Kawabata2019Symmetry,
  title = {Symmetry and Topology in Non-Hermitian Physics},
  author = {Kawabata, Kohei and Shiozaki, Ken and Ueda, Masahito and Sato, Masatoshi},
  journal = {Phys. Rev. X},
  volume = {9},
  issue = {4},
  pages = {041015},
  numpages = {52},
  year = {2019},
  month = {Oct},
  publisher = {American Physical Society},
  doi = {10.1103/PhysRevX.9.041015},
  url = {https://link.aps.org/doi/10.1103/PhysRevX.9.041015}
}

@article{Kawabata2019Classification,
  title = {Classification of Exceptional Points and Non-Hermitian Topological Semimetals},
  author = {Kawabata, Kohei and Bessho, Takumi and Sato, Masatoshi},
  journal = {Phys. Rev. Lett.},
  volume = {123},
  issue = {6},
  pages = {066405},
  numpages = {7},
  year = {2019},
  month = {Aug},
  publisher = {American Physical Society},
  doi = {10.1103/PhysRevLett.123.066405},
  url = {https://link.aps.org/doi/10.1103/PhysRevLett.123.066405}
}

@article{Kotz2023Topological,
  title = {Topological classification of non-Hermitian Hamiltonians with frequency dependence},
  author = {Kotz, Maximilian and Timm, Carsten},
  journal = {Phys. Rev. Res.},
  volume = {5},
  issue = {3},
  pages = {033043},
  numpages = {19},
  year = {2023},
  month = {Jul},
  publisher = {American Physical Society},
  doi = {10.1103/PhysRevResearch.5.033043},
  url = {https://link.aps.org/doi/10.1103/PhysRevResearch.5.033043}
}

@article{Xiao2026Symmetry,
  title = {Symmetry and topology of monitored quantum dynamics},
  author = {Xiao, Zhenyu and Kawabata, Kohei},
  journal = {Phys. Rev. B},
  volume = {113},
  issue = {13},
  pages = {134307},
  numpages = {16},
  year = {2026},
  month = {Apr},
  publisher = {American Physical Society},
  doi = {10.1103/1q83-jcbw},
  url = {https://link.aps.org/doi/10.1103/1q83-jcbw}
}

@article{Hatano1996Localization,
  title = {Localization Transitions in Non-Hermitian Quantum Mechanics},
  author = {Hatano, Naomichi and Nelson, David R.},
  journal = {Phys. Rev. Lett.},
  volume = {77},
  issue = {3},
  pages = {570--573},
  numpages = {0},
  year = {1996},
  month = {Jul},
  publisher = {American Physical Society},
  doi = {10.1103/PhysRevLett.77.570},
  url = {https://link.aps.org/doi/10.1103/PhysRevLett.77.570}
}

@article{Mandal2021Symmetry,
  author  = {I. Mandal and E. J. Bergholtz},
  title   = {Symmetry and Higher-Order Exceptional Points},
  journal = {Phys. Rev. Lett.},
  volume  = {127},
  pages   = {186601},
  year    = {2021},
  doi     = {10.1103/PhysRevLett.127.186601}
}

@article{LeongRoy2025AmplifiedMagneticCatalysis,
  author        = {Leong, Christopher A. and Roy, Bitan},
  title         = {Amplified magnetic catalysis in non-{H}ermitian Euclidean and hyperbolic {D}irac liquids},
  journal       = {arXiv preprint},
  year          = {2025},
  eprint        = {2510.02304},
  archivePrefix = {arXiv},
  primaryClass  = {cond-mat.str-el},
  doi           = {10.48550/arXiv.2510.02304}
}

@article{Minganti2019PRA,
  author  = {Minganti, Fabrizio and Miranowicz, Adam and
             Chhajlany, Ravindra W. and Nori, Franco},
  title   = {Quantum exceptional points of non-Hermitian Hamiltonians and
             Liouvillians: The effects of quantum jumps},
  journal = {Phys. Rev. A},
  volume  = {100},
  pages   = {062131},
  year    = {2019},
  doi     = {10.1103/PhysRevA.100.062131}
}

@article{Davies1974CMP,
  author  = {Davies, E. B.},
  title   = {Markovian master equations},
  journal = {Commun. Math. Phys.},
  volume  = {39},
  pages   = {91--110},
  year    = {1974},
  doi     = {10.1007/BF01608389}
}

@article{Kossakowski1977CMP,
  author  = {Kossakowski, Andrzej and Frigerio, Alberto and
             Gorini, Vittorio and Verri, Maurizio},
  title   = {Quantum detailed balance and {KMS} condition},
  journal = {Commun. Math. Phys.},
  volume  = {57},
  pages   = {97--110},
  year    = {1977},
  doi     = {10.1007/BF01625769}
}

@article{LeongRoy2026PRB,
  author  = {Leong, Christopher A. and Roy, Bitan},
  title   = {Non-{H}ermitian catalysis of spontaneous symmetry breaking on Euclidean and hyperbolic lattices},
  journal = {Phys. Rev. B},
  volume  = {113},
  pages   = {155152},
  year    = {2026},
  doi     = {10.1103/3l8x-ffmn}
}

@article{MurshedRoy2025SciPost,
  author  = {Murshed, Sk Asrap and Roy, Bitan},
  title   = {Yukawa-{L}orentz symmetry of interacting non-{H}ermitian birefringent {D}irac fermions},
  journal = {SciPost Phys.},
  volume  = {18},
  pages   = {073},
  year    = {2025},
  doi     = {10.21468/SciPostPhys.18.2.073}
}

@article{MurshedRoy2024JHEP,
  author  = {Murshed, Sk Asrap and Roy, Bitan},
  title   = {Quantum electrodynamics of non-{H}ermitian {D}irac fermions},
  journal = {J. High Energy Phys.},
  volume  = {01},
  pages   = {143},
  year    = {2024},
  doi     = {10.1007/JHEP01(2024)143}
}

\end{document}